\documentclass{article}
\usepackage{amsfonts,amssymb,amsmath, graphicx}
\usepackage{theorem}
\usepackage{caption}
\usepackage{cite}
\usepackage{url}
\usepackage{vmargin}
\usepackage{enumerate}
\usepackage{tikz}
\usepackage{todonotes} 
\usepackage{color}
\usetikzlibrary{arrows}
\usetikzlibrary{shapes.geometric}
\tikzstyle{block}=[draw opacity=0.7,line width=1.4cm]
\usetikzlibrary{positioning,arrows,chains,matrix,scopes,fit,decorations.markings,decorations,decorations.pathreplacing}
\tikzset{
  on each segment/.style={
    decorate,
    decoration={
      show path construction,
      moveto code={},
      lineto code={
        \path [#1]
        (\tikzinputsegmentfirst) -- (\tikzinputsegmentlast);
      },
      curveto code={
        \path [#1] (\tikzinputsegmentfirst)
        .. controls
        (\tikzinputsegmentsupporta) and (\tikzinputsegmentsupportb)
        ..
        (\tikzinputsegmentlast);
      },
      closepath code={
        \path [#1]
        (\tikzinputsegmentfirst) -- (\tikzinputsegmentlast);
      },
    },
  },
  mid arrow/.style={postaction={decorate,decoration={
        markings,
        mark=at position .4 with {\arrow[#1]{stealth}}
      }}},
}

\newtheorem{definition}{Definition}[section]

\newtheorem{proposition}[definition]{Proposition}

\newtheorem{theorem}[definition]{Theorem}

\newtheorem{lemma}[definition]{Lemma}
\newtheorem{rmk}[definition]{Remark}
\numberwithin{equation}{section}

\DeclareMathOperator{\tr}{tr}

\newcommand{\beq}{\begin{equation}}
\newcommand{\eeq}{\end{equation}}
\newcommand{\bea}{\begin{eqnarray}}
\newcommand{\eea}{\end{eqnarray}}
\newcommand{\beano}{\begin{eqnarray*}}
\newcommand{\eeano}{\end{eqnarray*}}
\newcommand{\bma}{\begin{pmatrix}}
\newcommand{\ema}{\end{pmatrix}}

            \def\cF{{\cal F}}
\def\cG{{\cal G}}            
            \def\cL{{\cal L}}

\def\ff{{\mathfrak f}}
\def\fg{{\mathfrak g}}

\newcommand{\ZZ}{{\mathbb Z}}

\newcommand{\wh}[1]{\widehat{#1}}
\newcommand{\wt}[1]{\widetilde{#1}}

\newcommand{\prf}{\underline{Proof}:\ }
\newcommand{\finprf}{\null \hfill {\rule{5pt}{5pt}}\indent}
\newcommand{\ie}{{\it i.e.}\ }
\newcommand{\cf}{{\it c.f.}\ }

\title{On consistency around a $3 \times 3\times 3$ cube and Q3 analogue  of \\ the lattice Boussinesq equation}
\author{Pengyu Sun, ~ Cheng Zhang, ~ Frank Nijhoff
 }
\date{\empty}

\begin{document}
\maketitle
\begin{abstract}In this paper, we present two new aspects of lattice Boussinesq (BSQ) equations. First, we show that the lattice potential BSQ (lpBSQ) equation defined on a nine-point square lattice admits a natural extension of three-dimensional consistency to a $3\times 3\times 3$ cube\textemdash a cubic sublattice consisting of $27$ vertices. This extends the standard notion of three-dimensional consistency (defined on an elementary $2\times 2\times 2$ vertex cube for quadrilateral equations) to the non-quadrilateral, nine-point setting. Second, we construct a new three-component system which is referred to as the {\em lattice BSQ-Q3 system},  serving as the BSQ analogue of the Q3($\delta$) equation in the Adler-Bobenko-Suris (ABS) classification. The construction relies on a gauge transformation between Lax pairs of lpBSQ with the parameter $\delta$ arising from a $GL_3$ action. In a degeneration form, the system yields a  $PGL_3$-invariant integrable lattice equation that generalises the  $PGL_2$-invariant Schwarzian BSQ equation.   
   \end{abstract}


\section{Introduction}

In the last two decades, the theory of integrable partial difference equations (P$\Delta$Es) on multi-dimensional lattices, or simply {\em integrable lattice equations}, has made great advances. In particular, the notion of {\em three-dimensional consistency} has emerged as one of the defining features for such equations\textemdash property that  a lattice equation or system can be consistently embedded into a higher-dimensional lattice \cite{NW, BS, Nij} (see also the monographs \cite{HJN, BS2}).

A major breakthrough was the classification result of Adler, Bobenko and Suris (ABS) which identified a list of affine-linear, 3D-consistent lattice quadrilateral equations\textemdash equations defined an elementary square lattice \cite{ABS, ABS2}. Most of ABS equations were known or related to known lattice Korteweg-de Vries-type (KdV-type) equations (see \cite{NC} for a review). The classification technique as well as the interrelations among the ABS equations exhibit rich structures in connection to classical invariant theory,  algebraic curves and B\"acklund transformations \cite{ABS2, AS, AJ}. Exact solutions of ABS equations were also extensively studied in a series of works \cite{AHN1, AHN, AN2, NAH, HZ2}.

In the ABS list, the Q3 equation holds a distinguished place, since the remaining ones, apart from the elliptic Q4 equation\cite{Adler
},  can be derived from it through certain degeneration processes  
\cite{AS, NAH}. The original form of Q3, first appeared in \cite{ABS},  reads
\begin{equation}
\label{eq:kdvq3abs}    
		\left(
		P-\frac{1}{P} \right)(\chi\widetilde{\chi}+\widetilde{\chi}\widehat{\widetilde{\chi}}) -\left(
		Q-\frac{1}{Q} \right)(\chi\widehat{\chi}+\wh{\chi}\widehat{\widetilde{\chi}}) = \left(
		\frac{P}{Q}-\frac{Q}{P} \right) \left(\widetilde{\chi}\widehat{\chi}+\chi\widehat{\widetilde{\chi}}+ \frac{\delta}{PQ} \right)\,,
            \end{equation}
            where $\chi:= \chi(n,m)$, $n,m \in \ZZ$, is the Q3 variable with its shifts in the lattice denoted by
            \begin{equation}
    \label{eq:th}          \wt{\chi} = \chi(n+1,m)\,,\quad               \wh{\chi} = \chi(n,m+1)\,,\quad \wh{\wt{\chi}} = \chi(n+1,m+1)\,, 
            \end{equation}
and $P, Q$ play the roles of lattice parameters associated to  respectively the  ~$\wt{~}$~ and ~$\wh{}$~  shifts. There is an extra fixed parameter  $\delta$ independent of the lattice directions in Q3. For $\delta=0$, Q3 is equivalent to the so-called  NQC equation \cite{NQC}.  

We will refer to \eqref{eq:kdvq3abs} as the lattice KdV-Q3 equation, and one of the main goals of this paper is to generalise KdV-Q3 to the next class of integrable lattice equations, \ie lattice Boussinesq-type (BSQ-type) equations, in the  {\em discrete  Gel’fand-Dikii hierarchy} \cite{GD}.

The prototype of lattice BSQ-type equations is the lattice potential BSQ (lpBSQ) equation, first given in \cite{DJM}. It reads
\begin{equation}\label{eq:dBSQ}
  \wh{\wh{\wt{\wt{w}}}}(\wh{\wh{\wt{w}}}-\wh{\wt{\wt{w}}})+w(\wh{w}-\wt{w})+(\wt{w}\wh{\wt{\wt{w}}}-\wh{w}\wh{\wh{\wt{w}}})=\frac{\alpha^3-\beta^3}{\wh{\wt{w}}-\wt{\wt{w}}}-\frac{\alpha^3-\beta^3}{\wh{\wh{w}}-\wh{\wt{w}}}\,,
\end{equation}where $w:=w(n,m)$, $n,m\in \ZZ$, is the lpBSQ variable with its shifts in the independent variables $n$ and $m$ denoted respectively by ~$\wt{~}$~ and ~$\wh{}$~ as in \eqref{eq:th}, and $\alpha$ and $\beta$ are respectively the lattice parameters associated to the  ~$\wt{~}$~ and ~$\wh{}$~  shifts. The lpBSQ equation is defined on a nine-point square lattice as shown in Figure~\ref{fig:nine-point-stencil}. It plays a similar role as the lpKdV equation for lattice KdV-type equations. Both lpKdV and lpBSQ can be derived from Hirota's bilinear Kadomtsev–Petviashvili (KP) equation through certain dimensional reductions \cite{HM}. A classification was also formulated for three-dimensional equations on octahedral lattices which contains lattice equations of KP-type \cite{ABS_KP}. 


\medskip
\begin{center}
    \begin{figure}[htb] 
    \centering
  \setlength{\unitlength}{.7mm}
\begin{picture}(160,60)(-60,0)
\put(0,0){\circle*{4}}
\put(-5,4){$w$} 
\put(0,30){\circle*{4}}
\put(-5,34){${{\wh{w}}}$}
\put(0,60){\circle*{4}}
\put(-5,64){${\widehat{\widehat{w}}}$}
\put(30,0){\circle*{4}}
\put(25,4){$ \widetilde{w}$}        
\put(30,30){\circle*{4}}
\put(25,34){${{\wh{\wt{w}}}}$}
\put(30,60){\circle*{4}}
\put(25,64){${{\wh{\wh{\wt{w}}}}}$}
\put(60,0){\circle*{4}}
\put(55,4){$\wt{\wt{w}}$}
\put(60,30){\circle*{4}}
\put(55,34){${{\wh{\wt{\wt{w}}}}}$}
\put(60,60){\circle*{4}}
\put(55,64){$\wh{\wh{\wt{\wt{w}}}}$}
\put(0,0){\line(0,1){30}}
\put(0,0){\line(1,0){30}}
\put(30,30){\line(-1,0){30}}
\put(30,30){\line(0,-1){30}}
\put(30,30){\line(1,0){30}}
\put(30,30){\line(0,1){30}}
\put(0,60){\line(1,0){30}}
\put(0,60){\line(0,-1){30}}
\put(60,0){\line(-1,0){30}}
\put(60,0){\line(0,1){30}}
\put(60,60){\line(-1,0){30}}
\put(60,60){\line(0,-1){30}}
\end{picture}
\caption{9-point configuration of the lattice BSQ type equations.}  \label{fig:nine-point-stencil}
\end{figure}
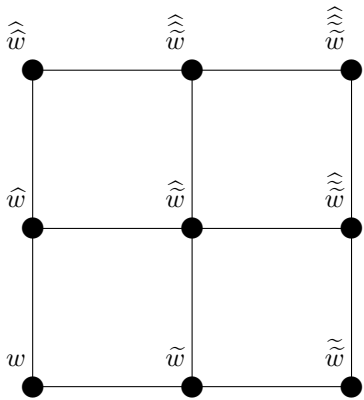
\end{center}


In contrast to the lattice KdV-type equations, for which a complete classification has been achieved,   the lattice BSQ-type equations remains significantly less developed. On the one hand, there is a lack of genuinely new examples.
A list of lattice BSQ-type equations was provided by Hietarinta \cite{H} (based on 3D consistency of a special form of multi-component system on a square lattice). All equations in the list can be identified with known systems derived from earlier constructions \cite{GD, TN2}, but some new parameters were found in some of the equations, which are referred to as {\em extended BSQ-type systems}. We will not consider these systems in the present paper. 
Analogues of Q-type BSQ as well as interrelations among the lattice BSQ equations organised around Q-type equations are missing.
In \cite{Nij-Q3}, a construction analogous to the KdV-Q3 equation was initiated for the BSQ case. However, it is not clear whether the procedure could lead to a closed-form equation, leaving the existence of a genuine lattice BSQ-Q3 equation as an open problem.
On the other hand, an integrability criterion, such as 3D consistency, is missing for the nine-point lpBSQ equation, since the standard notion of 3D consistency does not directly apply to  non-quadrilateral equations.

These limitations reflect a deeper gap of our understanding of BSQ-type equation: unlike the KdV (ABS) case, for which rich structures in connections to classical invariant theory, algebraic curves, and B\"acklund transformations (see, for instance, \cite{ABS2}) provide a unifying framework, no such comprehensive structure is yet known for the BSQ case. It seems that the development of new  lattice  BSQ-type equations\textemdash on par with the ABS list\textemdash requires new ideas and  technical concepts.

The aim of this paper is to address the above-mentioned problems for the lattice BSQ-type equations: the lack of a well-understood consistency property for nine-point lattice equations, and the absence of a proper analogue of the KdV-Q3 equation in the BSQ context. We achieve this goal through two principal advances. First, we establish that the lpBSQ equation exhibits a novel form of integrability: consistency around a $3\times 3\times 3$ cube (a $27$-vertex cubic lattice formed by $8$ elementary cubes). This property extends the standard notion of 3D consistency beyond the quadrilateral case. 
Second, inspired by the structure of the KdV-Q3 equation, we construct a lattice BSQ-Q3 system\textemdash a three-component system defined on an elementary square lattice satisfying  3D consistency.

The lattice BSQ-Q3 system we obtained can be written as follows: let $u,v,z$ denote the lattice BSQ-Q3 variables defined on a 2D lattice \begin{equation}
    u := u(n, m )\,,\quad v := v(n, m )\,, \quad z := z(n, m ) \,,\quad   n, m \in {\mathbb Z}  \,,\end{equation}
  and we employ the ~$\wt{~}$~, ~$\wh{~}$~ notation to denote shifts in $n$ and $m$ as in \eqref{eq:th}. Let $\alpha, \beta$ denote the lattice parameters associated respectively to ~$\wt{~}$~, ~$\wh{~}$~ shifts, and let $p,q$ be two extra parameters with $p\neq q$.   To simplify the expression, define the following set of parameters\footnote{The cubic power reflects the natural dispersion relation of BSQ-type equations. }
  \begin{equation}\label{sstt}
    s^3 = p^3 -\alpha^3\,,\quad \sigma^3 = q^3 -\alpha^3\,,\quad t^3 = p^3 -\beta^3\,,\quad \tau^3 = q^3 -\beta^3\,,
  \end{equation}
and also the quantities
  \begin{equation}
    \label{eq:fguv}
    f_u =\sigma^3 u -s^3  \wt{u}  \,, \quad f_v = \sigma^3 v  -  s^3 \wt{v}\,,\quad
    g_u =\tau^3  u -t^3 \wh{u}  \,,\quad  g_v =  \tau^3 v-t^3 \wh{v} \,.      
  \end{equation}
The lattce BSQ-Q3 system can be expressed as
\begin{subequations}
  \label{eq:fuvz}
  \begin{align}
    \label{eq:thuu}
    \wh{\wt{u}} = &\frac{ s^3 t^3 \left[ \wt{u} g_u -\wh{u} f_u +  r( \wh{v}-\wt{v})
                    \right]z +
                    s^3 t^3( \wt{u}g_v-\wh{u}f_v)+  \rho r^2 \Delta }{
                    s^3 t^3 \left[ (g_u-f_u)  z+ g_v - f_v \right]} \,,\\
    \label{eq:thv} \wh{\wt{v}} = &\frac{\left[s^3 t^3
         ( \sigma^3 \wh{v} (\wt{u} - u) - \tau^3 \wt{v}( \wh{u} -  u))
   +\rho r^2\Delta \right]
                   z+
                       s^3 t^3 (\wt{v} g_v - \wh{v} f_v  )         +r( \tau^3 f_u -\sigma^3 g_u ) \Delta      }{
s^3 t^3 \left[ (g_u-f_u)  z+ g_v -f_v \right]
                   }\,,\\
    \label{eq:tz}
                   \wt{z} =& \frac{\left[f_u f_v + r u f_v   +r^2\Delta \right]z  + f_v^2 +r (vf_v - f_u \Delta) }{\left[-f_u^2 -r(u f_u+f_v) -r^2v \right]z -f_u f_v - r uf_v  -r^2\Delta}\,, \\\label{eq:hz}
    \wh{z} =& \frac{\left[g_u g_v +r u g_v   +r^2\Delta \right]z  + g_v^2 +r(vg_v - g_u \Delta) }{\left[-g_u^2 -r(u g_u+g_v) -r^2v \right]z -g_u g_v - r ug_v  -r^2\Delta} \,,
  \end{align}
\end{subequations}
where
  \begin{equation}
 r =  p^3 - q^3\,,\quad \rho = \alpha^3 -\beta^3\,,\quad    \Delta  =  \frac{\sigma^{3n}\tau^{3m}}{s^{3n}t^{3m}}\delta \,, 
  \end{equation}
  with $\delta$ being a fixed parameter playing a role analogous to that in the lattice KdV-Q3 equation.
  The system \eqref{eq:fuvz} is non-autonomous defined on a square lattice, with \eqref{eq:tz} and \eqref{eq:hz} being understood as ``side equations'' (defined on two sides of the square).  All right-hand-side expressions in \eqref{eq:fuvz}  are in fractional linear forms in $z$.    
Lattice BSQ-Q3 is 3D-consistent and can be related to the lpBSQ equation \eqref{eq:dBSQ} via Miura-type transformation. Extra internal consistency as well as an autonomous version  of the above system will also  be discussed (see Section $4$). As $r=0$, the system reduces to a  $PGL_3$-invariant system generalising the well-known $PGL_2$-invariant Schwarzian BSQ equation.

This work contributes to a broader programme aimed at extending integrable lattice equations beyond the quadrilateral (KdV-type) setting. The main results presented here: the 3D consistency of the nine-point lpBSQ equation and the construction of a 3D-consistent lattice BSQ–Q3 system, provide concrete steps toward such objective. In particular, the search for a BSQ analogue of the elliptic Q4 equation remains one of the most distinguished open problems in this direction.

The paper is organised as follows. In Section $2$, we revisit the lpKdV equation and its associated lattice KdV-Q3 equation, providing a concise overview of the construction that motivates our approach to lattice BSQ-Q3. Section $3$ is devoted to the consistency of lpBSQ equation  around a $3\times 3\times 3$ cube. 
In Section $4$, we provide the derivation of the lattice BSQ-Q3$(\delta)$ system. We show its 3D consistency and study a reduced form which is  $PGL_3$-invariant.  Section $5$ contains concluding remarks.
\section{LpKdV and lattice KdV-Q3 equation}\label{sec:KdV}
This section first reviews some basic integrability aspects of the lpKdV equation such as its Lax pair and 3D consistency property. The main result of this section is to provide an alternative approach to deriving the lattice KdV-Q3 equation \eqref{eq:kdvq3abs}  through a ``discrete'' gauge transformation between two Lax pairs of lpKdV with different spectral parameters. This approach clarifies the origin of the $\delta$ term appearing in \eqref{eq:kdvq3abs} that arises from  a $GL_2$ action. This approach inspires us to extend lattice KdV-Q3 to the BSQ case  \eqref{eq:fuvz} which  will be the main subject of Section $4$.

\subsection{Lax pair and 3D consistency of LpKdV}
We recall some basic aspects of the lpKdV equation. We refer readers to \cite{NC,  BS, HJN} for details. 

Let $w$ denote the lpKdV variable that is a discrete field defined on a multi-dimensional lattice. In  three-dimensional case,   $w:=w(n,m,\ell)$, $n,m,\ell \in \mathbb Z$. We follow the convention that shifts  in the independent variables $n$, $m$ and $l$  are respectively denoted  by  ~$\wt{~}$~, ~$\wh{~}$~, ~$\bar{~}$~, \ie
\begin{equation}\label{eq:thb}
     \wt{w} = w(n+1,m,\ell)\,,\quad \wh{w} = w(n,m+1,\ell)\,,\quad \overline{w} = w(n,m,\ell+1)\,, \quad \dots\,. 
\end{equation}

The lpKdV equation is defined on an elementary square. In the $(n ,m) $ plane, it reads
\begin{equation} \label{eq:H1}
(  \widetilde{w} -\widehat{w})(\widehat{\widetilde{w}}-w) =\alpha^2-\beta^2\, ,
\end{equation}
where $\alpha$ and $\beta$ are lattice parameters associated respectively to the ~$\wt{~}$~,  ~$\wh{~}$~ shifts. 
LpKdV  \eqref{eq:H1} is a result of the compatibility of the linear systems 
\begin{equation}\label{eq:KDVUV}
    U_p\Phi_p =\wt{\Phi}_p\,,\quad     V_p\Phi_p =\wh{\Phi}_p\,,
\end{equation}
where $U_p$ and $V_p$ are $2\times 2$ Lax matrices in the forms
\begin{equation}\label{laxmah1}
	U_p =  \bma -\widetilde{w} & 1 \\  p^2-\alpha^2  -w\,\widetilde{w}& w \ema \,, \quad V_p = \bma -\widehat{w} & 1 \\  p^2-\beta^2  -w\,\widehat{w}& w \ema \,. 
\end{equation}
Here ${\Phi}_p$ is understood as a fundamental solution, \ie nondegenerate $2\times 2 $ matrix-valued solution, of the linear system \eqref{eq:KDVUV}, and $p^2$ serves as the spectral parameter (see Remark \ref{sec:21} below). Since     $\det U_p = \alpha^2 - p^2$ and $ \det V_p = \beta^2 - p^2$, we normalize  $\Phi_p$   as 
\begin{equation}\label{eq:normphi}
    \det \Phi_p  = (\alpha^2 - p^2)^n(\beta^2 - p^2)^m(\gamma^2 - p^2)^\ell \,, 
\end{equation}
where the term $(\gamma^2 - p^2)^\ell $ is a result of  $W_p \Phi_p =\overline{\Phi}_p$ as the linear equation in the $\ell$ direction with $W_p$ defined similarly as $U_p$ by switching (~$\wt{~}~,  \alpha)$ to  (~$\bar{~~}~,  \gamma)$. 

\begin{figure}[htb]
	\centering
	\begin{tikzpicture}[scale=0.55, decoration={markings,mark=at position 0.55 with {\arrow{latex}}}]
		\tikzstyle{nod}= [circle, inner sep=0pt, fill=white, minimum size=7pt, draw]
		\tikzstyle{nod1}= [circle, inner sep=0pt, fill=black, minimum size=7pt, draw]
         \tikzstyle{nodd}= [diamond, fill=white, isosceles triangle stretches,shape border rotate=270,minimum size=6pt, draw]
   \tikzstyle{nodsw}= [rectangle, inner sep=0pt, fill=white, minimum size=8pt, draw]
		\def\lx{3}%
		\def\ly{1.22}%
		\def\lz{ (sqrt(\x*\x+\y*\y))}%
		\def\l{4}%
		\def\d{4}%
		\coordinate (u00) at (0,0);
		\coordinate (u10) at (\l,0);
		\coordinate (u01) at (\lx,\ly);
		\coordinate (u11) at (\l+\lx,\ly);
		\coordinate (v00) at (0,\d);
		\coordinate (v10) at (\l,\d);
		\coordinate (v01) at (\lx,\d+\ly);
		\coordinate (v11) at (\l+\lx,\d+\ly);
		\draw[-]  (u10)  
		-- (u11) ;
		\draw [dashed]  (u01)  -- (u11);
		\draw [dashed]  (u00)--node [above]{$\beta$}(u01) ;
		\draw[-] (v00) --  (v10) -- (v11) -- (v01) -- (v00);
		\draw[-] (u11) --  (v11);
		\draw[-] (u00) -- node [left]{$\gamma$}(v00);
		\draw[-] (u10) -- (v10);
		\draw[dashed] (u01) -- (v01);
		\coordinate (u011) at (1.5*\lx,1.5* \ly);
		\draw[dashed] (u01) -- (u011) ;
		\draw[dashed] (u01) -- (.66*\lx,\ly) ;
		\draw[-] (u00) -- (-0.5*\lx,-0.5* \ly);
		\draw[-] (u00) -- (-.33*\lx,0);
		\draw[-] (u10) -- (0.33*\lx+\l,0);
		\draw[-] (u10) -- (-.5*\lx+\l,-.5*\ly);
		\draw[-] (u11) -- (1.33*\lx+\l,\ly);
		\draw[-] (u11) -- (-.5*\lx+\l,-.5*\ly);
		\draw[-]  (u11)-- (1.5*\lx+\l,1.5* \ly);
		\draw[-] (u00) 
		-- node [below]{$\alpha$} (u10);
		\node[nod1] (v00) at (0,\d) [label=above: $\overline{w}$] {};
		\node[nodsw] (v10) at (\l,\d) [label=above: $\widetilde{\overline{w}}$] {};
		\node[nodsw] (v01) at (\lx,\d+\ly) [label=above: $\widehat{\overline{w}}$] {};
		\node[nodd] (v11) at (\l+\lx,\d+\ly) [label=above: $\widehat{\widetilde{\overline{w}}}$] {};
		\node[nod1] (u00) at (0,0) [label=below: $w$] {};
		\node[nod1] (u10) at (\l,0) [label=below: $\widetilde{w}$] {};
		\node[nod1] (u01) at  (\lx,\ly) [label=below: $\widehat{w}$] {};
		\node[nodsw] (u11) at (\l+\lx,\ly) [label=below: $\widehat{\widetilde{w}}$] {};
	\end{tikzpicture}
	\caption{3D consistency of lpKdV: the six faces of an elementary cube are governed by the lpKdV equations.  Having values of $(w, \wt{w}, \wh{w}, \overline{w})$ on the {\em black dots} as well as of the lattice parameters $(\alpha, \beta, \gamma)$ as initial data, one first computes  the values of $(\wh{\wt{w}}, \wt{\overline{w}}, \wh{\overline{w}})$ on the {\em white squares}, then has three ways to determine the value on the {\em white diamond}.  Consistency means that this  is well-posed. } \label{fig:cac}
\end{figure}
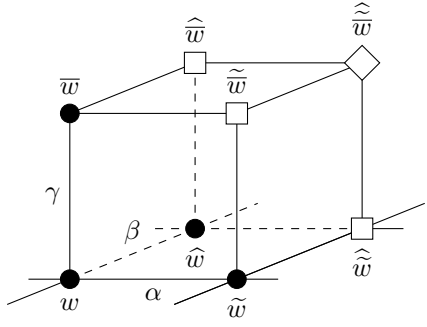

The lpKdV variable $w$ is consistent in a 3D lattice. This is an intrinsic  property of the lpKdV equation. On an elementary square lattice,  lpKdV \eqref{eq:H1} is affine-linear with respect to $w$ and its shifts, which implies that an initial-value problem with generic initial data on three vertices as well as the lattice parameters determines uniquely the  value on the fourth one. 
LpKdV \eqref{eq:H1} is said to be {\em 3D-consistent}, if it can be consistently posed as an initial-value problem on an elementary cube in a three-dimensional lattice. This is explained in Figure~\ref{fig:cac}.    Note that the above consistency property  is a common feature for all lattice KdV-type equations listed in the ABS classification \cite{ABS, ABS2}.


\begin{rmk}\label{sec:21}Let $(\phi_1, \phi_2)^\intercal$ be a $2$-vector-valued solution of the linear system \eqref{eq:KDVUV}. By elementary computations, one can show that $\phi_1$, $\phi_2$ obey 
\begin{subequations}\label{eq:scalarLin}
    \begin{align}
       \label{eq:2so1} \wt{\wt{\phi}}_1 + (\wt{\wt{w}} - w) \wt{\phi}_1 + \alpha^2 \phi_1 & = p^2 \phi_1  \\
        \label{eq:2so2}        \wh{\wh{\phi}}_1 + (\wh{\wh{w}} - w) \wh{\phi}_1 + \beta^2 \phi_1 & = p^2 \phi_1 
    \end{align}
    and 
    \begin{equation}
        \phi_2 = \wt{\phi}_1 + \wt{w}\, {\phi}_1 =  \wh{\phi}_1 + \wh{w}\, {\phi}_1  \,. 
    \end{equation}
\end{subequations}
The above linear system for $\phi_1$ can also be seen as Lax pair for lpKdV where $p^2$ serves as the spectral parameter in the spectral problems \eqref{eq:2so1} and \eqref{eq:2so2}.  
\end{rmk}

\subsection{Lattice KdV-Q3 equation and origin of  $\delta$}\label{sec:22}
This section aims to provide a new constructive approach to the lattice KdV-Q3 equation using the Lax pairs of lpKdV, alternative to the constructions in  \cite{AHN1,Nij-Q3} (see also Section $9.6$ of \cite{HJN}) which are based on the {\em direct linearization method} via  Cauchy-type integral representations. The advantage of the present approach is that it relies only on the algebraic structure of the Lax pairs, without invoking global solution representations. Note that an inverse scatterng transform for KdV-Q3 was also developed in \cite{Butler}. 

Let $u$ be the KdV-Q3 variable that is a discrete field on a 3D lattice, \ie $u:=u(n,m,\ell)$. In our setting, the lattice KdV-Q3 equation in the $(n,m)$ plane reads (the convention \eqref{eq:thb} is followed)
\begin{equation}\label{eq:Q3d}
   (\alpha^2 - q^2)u\wh{u}+   (\alpha^2 - p^2)\wt{u}\wh{\wt{u}}-   (\beta^2 - q^2)u\wt{u}-   (\beta^2 - p^2)\wh{u}\wh{\wt{u}} = (\alpha^2 - \beta^2)(\wt{u} \wh{u} +u\wh{\wt{u}} + \Delta)\,,
\end{equation}
where  $p\neq q$, and 
\begin{equation}
    \Delta = \frac{(\alpha^2 - q^2)^n(\beta^2 - q^2)^m(\gamma^2 - q^2)^\ell  }{(\alpha^2 - p^2)^n(\beta^2 - p^2)^m (\gamma^2 - p^2)^\ell} \,\delta  \,,
\end{equation}
with $\delta$ being an arbitrary constant. Besides the lattice parameters $\alpha, \beta, \gamma$, it contains  $p, q$ and $\delta$ as extra parameters. 
The introduction 
of the ``new'' variable $u$ obeying \eqref{eq:Q3d}    is rather subtle and forms the key to the 
construction of the KdV-Q3 equation.
Before going into details of this construction, we first state some interesting features of  \eqref{eq:Q3d}. 
\begin{itemize}
    \item \eqref{eq:Q3d} is 3D-consistent. 
    \item \eqref{eq:Q3d} is non-autonomous, one can put it in an autonomous form. For simplicity, let $\ell = 0$ and
    \begin{equation}
        s^2 = \alpha^2 - p^2\,,\quad         \mu^2 = \alpha^2 - q^2\,,\quad         t^2 = \beta^2 - p^2\,,\quad        \nu^2 = \beta^2 - q^2\,,
    \end{equation}
    where the branches of $s,t, \mu,\nu$ are fixed such that standard exponent rules apply. Let 
    \begin{equation}
        \chi = u\, \frac{\mu^n\,\nu^m}{s^n\,t^m}\, ,
    \end{equation}
    then $\chi$ obeys 
    \begin{equation}
        s\, \mu(\chi \wh{\chi} + \wt{\chi}\wh{\wt{\chi}} ) - t \,\nu (\chi \wt{\chi}+ \wh{\chi}\wh{\wt{\chi}} ) = (\alpha^2-\beta^2)(\wt{\chi}  \wh{\chi}+\chi \wh{\wt{\chi}}+ \frac{s\,t}{\mu\,\nu}\delta)\,. 
    \end{equation}Using the parametrization $P = \mu/s, Q = \nu/t$,  it can be transformed into  \eqref{eq:kdvq3abs}. 
      \item   Lattice KdV-Q3  is the ``second
top" equation in the ABS classification, as the remaining ones (apart from Q4) can be derived from it through
certain degeneration processes \cite{AS, NAH}. 
    \item  In the case $\delta = 0$ and $q = p $, \eqref{eq:Q3d} is reduced to the lattice Schwarzian KdV equation \cite{Weiss, NC}
    \begin{equation}
        \label{eq:lskdv}
        \frac{(u -\wh{u})(\wt{u}-\wh{\wt{u}})}{(u -\wt{u})(\wh{u}-\wh{\wt{u}})} =  \frac{\alpha^2 - p^2}{\alpha^2 - p^2}\,,
    \end{equation}
    which is invariant under an action of simultaneous fractional linear transformations on $u$ and its shifts. In other words, it is $PGL_2$-invariant.
  
\end{itemize}  

The derivation of \eqref{eq:Q3d} is split into the following steps.  For simplicity, fix $\ell = 0$.
The key idea is to relate two copies of the lpKdV Lax system with different spectral parameters via a gauge matrix, and the KdV-Q3 variable $u$ will be defined as the trace of the gauge matrix.

\begin{description}
    \item[Step 1: discrete gauge transformation.] Let $U_q$ and $ V_q$ be another pair of Lax matrices by switching the spectral parameter $p$ to $q$,  \ie  $U_q = U_p\vert_{p\to q}$ and $     V_q = V_p\vert_{p\to q}$ with $p\neq q$. Let $\Phi_q$ be the fundamental solution of 
\begin{equation}
  \label{eq:KDVUV1}
    U_q\Phi_q=\wt{\Phi}_q\,,\quad     V_q\Phi_q =\wh{\Phi}_q\,, 
\end{equation}
normalized similarly as \eqref{eq:normphi}. Clearly, compatibility condition of \eqref{eq:KDVUV1} again leads to lpKdV \eqref{eq:H1}. Let $G$ be a gauge matrix connecting $\Phi_p$ and $\Phi_q$ such as
\begin{equation}\label{eq:gkdvG}
    G \,\Phi_p = \Phi_q\,,
\end{equation}
then the pairs $U_p, V_p$ and $U_q, V_q$ are connected by 
    \begin{equation}\label{eq:gkdv}
        \wt{G}\, U_p = U_q\,G\,,\quad         \wh{G}\, V_p = V_q\,G\,. 
    \end{equation}
Due to the normalization \eqref{eq:normphi}, 
\begin{equation}\label{eq:gkdv1}
    \det G\vert_{\ell = 0} =  \frac{(\alpha^2 - q^2)^n(\beta^2 - q^2)^m  }{(\alpha^2 - p^2)^n(\beta^2 - p^2)^m } \,. 
\end{equation}

    \item[Step 2: assignment of the variable.] The lattice KdV-Q3 variable $u$ is defined as 
\begin{equation}\label{eq:utrg}
    u = \text{tr}(G) = g_{11}+g_{22}\,,
\end{equation}
where $g_{ij}= (G)_{ij} $ which is the $ij$-entry of $G$. In view of \eqref{eq:gkdvG}, one has
\begin{equation}
    u =  \frac{\begin{vmatrix}
        \phi_{q,11} & \phi_{p,12} \\\phi_{q,21} & \phi_{p,22} 
    \end{vmatrix} - \begin{vmatrix}
        \phi_{q,12} & \phi_{p,11} \\\phi_{q,22} & \phi_{p,21} 
    \end{vmatrix} }{(\alpha^2 - p^2)^n(\beta^2 - p^2)^m}\,,
\end{equation}
where $\phi_{p,ij} $ and $\phi_{q,ij}$ are respectively the $ij$-entry of $\Phi_p$ and $\Phi_q$, and $|\cdot|$ stands for  determinant.  
\item[Step 3: shifts of $u$.] Take (\ref{eq:gkdv}-\ref{eq:utrg}) as the governing systems: on the one hand, solving the linear systems \eqref{eq:gkdv} one can get ~$\wt{~}$~ and ~$\wh{~}$~ shifts of $g_{ij}$ as expressions of the remaining variables; on the other, using \eqref{eq:gkdv1} and \eqref{eq:utrg} one can get $g_{11}, g_{21}$ as expressions of $g_{12}$, $g_{22}$ and $u$. This allows to compute 
\begin{equation}
    \wt{u} =  \wt{g}_{11}+\wt{g}_{22}\,,\quad     \wh{u} =  \wh{g}_{11}+\wh{g}_{22}\,. 
\end{equation}
Precisely, one has
\begin{subequations}\label{eq:thu}
    \begin{align}\label{eq:thu1}
        \wt{u} =& \frac{(\alpha^2-q^2) u +(p^2-q^2)(g_{12}\,\wt{w} -g_{22})}{\alpha^2-p^2}\,, \\
    \label{eq:thu2}    \wh{u} = &\frac{(\beta^2-q^2) u +(p^2-q^2)(g_{12}\,\wh{w} -g_{22})}{\beta^2-p^2}\,.
    \end{align}
\end{subequations}

\item[Step 4: Miura-type transformation.] 
Consider \eqref{eq:thu}, one can express $\wt{w},\wh{w}$ in terms of $\wt{u},\wh{u}$ and $ g_{12}, g_{22}$. One can get 
\begin{equation}
    \label{eq:mkdv1} \wt{w}-\wh{w} = \frac{p^2 (\wh{u}-\wt{u})+\alpha^2(\wt{u}- u )-\beta^2(\wh{u}-u )}{g_{12}(p^2-q^2)} \,.
\end{equation}

Take ~$\wh{~}$~ shift of $\wt{u}$ following \eqref{eq:thu1} (one uses $\wh{\wt{u}}$ and $\wh{\wt{w}}$ to  denote respectively ~$\wh{~}$~ shift of $\wt{u}$ and $\wt{w}$). Using the substitutions explained in Step 3, one can get an expression of $\wh{\wt{w}}$ of  the remaining variables. It follows from direct computations that
    \begin{equation}
\label{eq:miura2}    \wh{\wt{w}}  -w  = {g_{12}}\, 
    \frac{(p^2-q^2)(\beta^2-p^2)(     q^2 u     - p^2 \wh{\wt{u}} +\beta^2(\wh{u}-{u}) +\alpha^2(\wh{\wt{u}}-\wh{u}  )  ) }{
   (\beta^2-p^2) (\wh{u}-u)(q^2 u -p^2 \wh{u}  + \beta^2 (\wh{u} - u) ) +(p^2 - q^2)^2 \det G
    }       \,.
    \end{equation}
The formulae \eqref{eq:mkdv1} and \eqref{eq:miura2} can be understood as Miura-type transformation transforming lpKdV to  lattice KdV-Q3: the differences of the lpKdV variable $w$ and its shifts are rational expressions of lattice KdV-Q3 variable $u$ and its shifts. Inserting them into lpKdV \eqref{eq:H1},  one gets \eqref{eq:Q3d} with $\delta =(\alpha^2 -p^2)(\beta^2 -p^2)$. 

There are some extra consistencies needed to be clarified. Similarly, taking ~$\wt{~}$~ shift of $\wh{u}$ using \eqref{eq:thu2} (one uses $\wt{\wh{u}}$ and $\wt{\wh{w}}$ to  denote respectively ~$\wt{~}$~ shift of $\wh{u}$ and $\wh{w}$), one can get 
\begin{equation}
    \label{eq:miura3} \wt{\wh{w}}  -w  = {g_{12}}\, 
    \frac{(p^2-q^2)(\alpha^2-p^2)(     q^2 u     - p^2 \wt{\wh{u}} +\alpha^2(\wt{u}-{u}) +\beta^2(\wt{\wh{u}}-\wt{u}  )  ) }{
   (\alpha^2-p^2) (\wt{u}-u)(q^2 u -p^2 \wt{u}  + \alpha^2 (\wt{u} - u))  +(p^2 - q^2)^2 \det G
    }       \,,
\end{equation}
which is of the same nature as \eqref{eq:miura2}. This allows to get the same lattice KdV-Q3 equation as in Step 4 by setting $\wt{\wh{w}}=\wh{\wt{w}}$ and  $\wt{\wh{u}}=\wh{\wt{u}}$. Moreover, eliminating $\wh{\wt{u}}$ in \eqref{eq:miura2} and \eqref{eq:miura3} provides an expression of $ \wh{\wt{w}}  -w $, which together with \eqref{eq:mkdv1} is consistent with lpKdV \eqref{eq:H1}.

\item[Step 5: origin of the $\delta$ term.]  Under the action of a $2\times 2$ constant matrix $M$ (which is not necessarily nondegenerate) on \eqref{eq:KDVUV1} as
  \begin{equation}\label{eq:kdvM}
      \Phi_q \mapsto \Phi_q^M =  \Phi_q\,M\,.
  \end{equation}   
$\Phi_q^M$ is again a solution of \eqref{eq:KDVUV1}. Using 
$  G\, \Phi_p  = \Phi_q^M$ and  $u = \tr G$, the above derivations still hold. This action induces 
\begin{equation}
    \det G  = \det M \, \frac{(\alpha^2 - q^2)^n(\beta^2 - q^2)^m }{(\alpha^2 - p^2)^n(\beta^2 - p^2)^m}\,,  
\end{equation}and setting $\delta  = \det M (\alpha^2 - p^2) (\beta^2 - p^2)$, one gets precisely \eqref{eq:Q3d}. Therefore, when $\delta \neq 0$, the $\delta$ term  is a result of a  $GL_2$ action.  Moreover, $u$ can be expressed as (see also \cite{AHN1, NAH, WZZZ})
\begin{equation}
    \frac{ m_{11} \begin{vmatrix}
     \phi_{q,11} &\phi_{p,12} \\
     \phi_{q,21} & \phi_{p,22}
   \end{vmatrix} - m_{12} \begin{vmatrix}
       \phi_{q,11} &\phi_{p,11} \\
     \phi_{q,21} & \phi_{p,21}
   \end{vmatrix} + m_{21} \begin{vmatrix}
          \phi_{q,12} &\phi_{p,12} \\
     \phi_{q,22} & \phi_{p,22}
   \end{vmatrix} - m_{22} \begin{vmatrix}
          \phi_{q,12} &\phi_{p,11} \\
     \phi_{q,22} & \phi_{p,21}
   \end{vmatrix} }{(\alpha^2 - p^2)^n(\beta^2 - p^2)^m}\,,
\end{equation}
where $|\cdot|$ stands for determinant, and $\phi_{p,ij} $, $\phi_{q,ij}$ and $m_{ij}$ are respectively the $ij$-entry of $\Phi_p$, $\Phi_q$ and $M$.
\end{description}

\begin{rmk}\label{rmk:gauge}
In the continuous theory of KdV-type equations, Miura-type transformations and related gauge constructions relating KdV and KdV-type equations were well developed within the framework of $SL_2$ loop group factorization  \cite{GM} (see also early classification results \cite{DS}). Our approach could be interpreted as a ``discrete'' realisation of the loop group factorization method. The present approach not only clarifies the algebraic origin of $\delta$ as induced by a $GL_2$ action, but also lays the foundation for higher-rank generalisations as demonstrated in Section $4$.

  \end{rmk}


\section{LpBSQ and consistency around a $3\times 3\times 3$ cube}\label{sec:3}
                We review two versions of the lpBSQ equation: one as a three-component system defined on an elementary square lattice; the other as a single-component nine-point equation defined on a (nine points) $3\times 3$-vertex square lattice. We show that the nine-point lpBSQ equation  exhibits a novel integrability property: consistency around a $3\times 3 \times 3$ cube.                   
\subsection{Lax pair and  consistency around a cube}
The lpBSQ equation, as a higher rank analogue of the lpKdV equation in the Gel'fand-Dikii hierarchy was first introduced in \cite{GD}. Here, we recall some basic results of lpBSQ, and collect notions and notations needed in the rest of the paper. We refer readers to \cite{GD, TN2,  HZ3, W} for details.

Consider the following first-order linear system 
\begin{equation}\label{eq:laxUV}
\widetilde{\Phi}_p = U_p \,\Phi_p\, \quad \widehat{\Phi}_p = V_p\, \Phi_p\,,
\end{equation}
where $U_p$ and $V_p$ are $3\times 3$ Lax matrices in the forms 
\begin{equation}
   U_p = \bma -\widetilde{w} & 1 & 0\\ -\widetilde{x} & 0 & 1\\  p^3-\alpha^3 + y\,\widetilde{w} -w\, \widetilde{x} & -y & w   \ema\,,\quad   V_p = \bma -\widehat{w} & 1 & 0\\ -\widehat{x} & 0 & 1\\  p^3 -\beta^3 + y\,\widehat{w} -w\, \widehat{x} & -y & w   \ema\,.
\end{equation}
Here $\alpha$ and $\beta$ are the lattice parameters  associated respectively to the ~$\widetilde{~}$~ and ~$\widehat{~}$~ shifts, and $p$ is related to the spectral parameter (see Remark~\ref{rmk:31} below). The lattice BSQ variables $w, x, y$ are defined on a multi-dimensional lattice.  In  three-dimensional case, 
\begin{equation}
    w:=w(n,m,\ell)\,, \quad x:=x(n,m,\ell)\,, \quad y :=y(n,m,\ell)\,, \quad n,m,\ell \in \ZZ\,,
\end{equation}and  shifts in  $n,m,\ell$ are respectively denoted  by ~$\wt{~}$~,~$\wh{~}$~,~$\bar{~}$~. For instance, 
\begin{equation}\label{eq:wnml}
  \widetilde{w}=w(n+1,m,\ell)\,,\quad \widehat{w}=w(n,m+1,\ell) \,,\quad \overline{w}=w(n,m,\ell+1)\,. 
\end{equation}
One also has $\overline{\Phi}_p = W_p \,\Phi_p$ with $W_p$ defined similarly as $U_p$ by switching  ($\wt{~},  \alpha)$ to  ($\bar{~~},  \gamma)$. 
Since
\begin{equation}
    \det U_p = p^3-\alpha^3 \,,\quad \det V_p =p^3 - \beta^3 \,, \quad \det W_p = p^3-\gamma^3 \,,
\end{equation}
the fundamental solution $\Phi_p$ is  normalized as 
\begin{equation}\label{eq:bsqn}
    \det \Phi_p =  (p^3-\alpha^3)^n (p^3 - \beta^3)^m  (p^3-\gamma^3)^\ell \,.
\end{equation}
 The compatibility of \eqref{eq:laxUV}, \ie $ \widetilde{ V}_p\, U_p = \widehat{U}_p \, V_p$, 
yields the following three-component system
\begin{subequations}\label{eq:3csyst}
\begin{align} \label{eq:3csyst1} \widehat{\widetilde{w}}  = &\frac{\widehat{x} -\widetilde{x} }{\widehat{w} -\widetilde{w} } \,,\\   \label{eq:3csyst2}w  = &\frac{\widehat{y} -\widetilde{y} }{\widehat{w} -\widetilde{w} } \,,\\  \label{eq:3csyst3}\widehat{\widetilde{x}} =&w  \widehat{\widetilde{w}} -y + \frac{\beta^3-\alpha^3}{\widehat{w} -\widetilde{w}} \,, 
\end{align}
 defined on   an elementary square lattice in the $(n,m)$ plane. Similar equations in other planes can also be derived using $
     \widetilde{ W}_p\, U_p = \overline{U}_p \, W_p$ and $
     \overline{ V}_p\, W_p = \widehat{W}_p \, V_p$.    
\end{subequations}
\begin{proposition}
This system \eqref{eq:3csyst} is consistent around a cube.    
\end{proposition}
To check its three-dimensional consistency, one needs to transform it into an {\em evolutionary form}. 
\begin{lemma} 
Let $\kappa$ be an arbitrary function in the form
\begin{equation}
    \kappa: = \kappa(n+m+l)\,,
\end{equation}then the system \eqref{eq:3csyst} is equivalent to 
\begin{subequations}\label{eq:3csystev}
    \begin{align}
\label{eq:3csystev1} x = &  w       \widetilde{w}-\widetilde{y}+ \widetilde{\kappa}\,,\\
\label{eq:3csystev2} x = &  w       \widehat{w}-\widehat{y}+ \widehat{\kappa}\,,\\
\label{eq:3csystev3}\widehat{\widetilde{x}} = &  w  \widehat{\widetilde{w}}  - y +\frac{\beta^3-\alpha^3}{\widehat{w} -\widetilde{w}} \,. 
    \end{align}
\end{subequations}
\end{lemma}
\prf the key observation is that combinations $   w       \widetilde{w}-\widetilde{y} -x$ and $w       \widehat{w}-\widehat{y} -x $ are constant along the characteristic lines $n+m = \text{const}$, which allows us to introduce the auxiliary function $\kappa$.

\eqref{eq:3csystev} $\implies$ \eqref{eq:3csyst}: \eqref{eq:3csyst3} and \eqref{eq:3csystev3} coincide. By definition, $ \widehat{\kappa} =\widetilde{\kappa}$, then the subtraction of \eqref{eq:3csystev1} and \eqref{eq:3csystev2} leads to  \eqref{eq:3csyst2}; the  subtraction of  ~$\widehat{~}$~ shift of \eqref{eq:3csystev1} and  ~$\widetilde{~}$~ shift of \eqref{eq:3csystev2} leads to \eqref{eq:3csyst1}.

\eqref{eq:3csyst} $\implies$ \eqref{eq:3csystev}:  we present  \eqref{eq:3csyst1} and \eqref{eq:3csyst2} in the forms
\begin{subequations}
    \begin{align}
\label{eq:wxy1} \widehat{w} \widehat{\widetilde{w}}-\widetilde{w}\widehat{\widetilde{w}}  = &\widehat{x} -\widetilde{x} \,,\\ 
\label{eq:wxy2} \widetilde{w} \wh{\widetilde{w}} - \widetilde{w}\wt{\widetilde{w}}  = &\widehat{\widetilde{y}} -\widetilde{\widetilde{y}} \,,\\ 
\label{eq:wxy3} \widehat{w} \widehat{\widehat{w}} -\widehat{w}\widehat{\widetilde{w}}  = &\widehat{\widehat{y}} -\wh{\widetilde{y}} \,,
\end{align}
\end{subequations}
where the last two equations are results of   \eqref{eq:3csyst2} under  ~$\widetilde{~}$~ and ~$\widehat{~}$~  shifts respectively. Adding \eqref{eq:wxy1} with \eqref{eq:wxy2} and \eqref{eq:wxy3}, one gets
\begin{subequations}
    \begin{align}
       ( ~\wh{~}~ - ~\wt{~}~) (w\wt{w} -x -\wt{y}) &= 0\,,\\
              ( ~\wh{~}~ - ~\wt{~}~) (w\wh{w} -x -\wh{y})&= 0\,,
    \end{align}
\end{subequations}
where  ~$\wh{~}$~  and  ~$\wt{~}$~ are understood as shift operators.  This implies that 
\begin{equation}
    w\wt{w} -x -\wt{y} = \kappa_1\,,\quad    w\wh{w} -x -\wh{y} = \kappa_2\,,
\end{equation}
where $\kappa_{j}$, $j=1,2$, is a solution of $( ~\wh{~}~ - ~\wt{~}~)\kappa_j =0$. Then,  $\kappa_{j}$ is in form of $\kappa$, due to \eqref{eq:3csyst1} and \eqref{eq:3csyst2}, $\kappa_1 = \kappa_2$, $\wh{\kappa}_1 = \wt{\kappa}_2$, which means $\kappa_1 \equiv \kappa_2$. Setting $\kappa_1 = \kappa_2 =\kappa $ concludes the proof. 
\finprf
\medskip

The system \eqref{eq:3csystev} is evolutionary in the following sense: solving the system involving  ~$\widehat{~}$~ shift of \eqref{eq:3csystev1} and  ~$\widetilde{~}$~ shift of \eqref{eq:3csystev2}, one obtains
\begin{equation}\label{eq:thbwxy}
     \widehat{\widetilde{w}}  = \frac{\widehat{x} -\widetilde{x} }{\widehat{w} -\widetilde{w} } \,,\quad  \wh{\wt{x}} = \frac{w(\wh{x}-\wt{x})+\beta^3-\alpha^3}{\wh{w}-\wt{w}}-y\,,\quad \wh{\wt{y}} = \frac{\wt{w}\wh{x}- \wh{w}\wt{x}}{\wh{w}-\wt{w}}- \widetilde{\kappa}\,. 
\end{equation}
Therefore,  on an elementary square lattice,  having $(w,\wt{w}, \wh{w},x,\wt{x}, \wh{x},y)$  as well as thee lattice parameters $\alpha, \beta$ as initial values, one can first obtain $\wt{y}$ and $\wh{y}$ using \eqref{eq:3csystev1} and \eqref{eq:3csystev2}, then determine $(  \widehat{\widetilde{w}} ,   \widehat{\widetilde{x}},   \widehat{\widetilde{y}}  )$. 
3D consistency property can be directly checked using \eqref{eq:thbwxy}.

\begin{rmk}\label{rmk:31}
 Using $(\phi_1,\phi_2,\phi_3)^\intercal$ as a solution of \eqref{eq:laxUV},  one can transform \eqref{eq:laxUV} into the following linear systems
\begin{subequations}
    \begin{align}
    \label{eq:tttvp}
\widetilde{\widetilde{\widetilde{  \phi}}}_1+ ( \widetilde{\widetilde{\widetilde{  w}}} -w)\widetilde{\widetilde{  \phi}}_1 +(\widetilde{\widetilde{  x}}+y -w\widetilde{\widetilde{w}}) \widetilde{\phi}_1  + \alpha^3 \phi_1 =& p^3 \phi_1\,, \\
\wh{\wh{\wh{  \phi}}}_1+ (\wh{\wh{\wh{w}}}-w)\wh{\wh{  \phi}}_1 +( \wh{\wh{  x}}+y -w\wh{\wh{w}}) \wh{\phi}_1  + \beta^3 \phi_1 = &p^3 \phi_1\,,
\end{align}
and
\begin{equation}
  \phi_2 = \widetilde{\phi}_1+\widetilde{w}\,\phi_1= \wh{\phi}_1+\wh{w}\,\phi_1\,, \quad \phi_3 = \widetilde{\phi}_2+\widetilde{x}\,\phi_1= \wh{\phi}_2+\wh{x}\,\phi_1\,. 
\end{equation}
\end{subequations}
The first two difference equations are third-order discrete spectral problems in the discrete Gel'fand-Dikii hierarchy \cite{GD} where $p^3$ is the spectral parameter. 
\end{rmk}

One can transform \eqref{eq:3csyst} into a single-component equation for $w$ as a nine-point equation defined on 
a $3\times 3 $-vertex square lattice (see Figure~\ref{fig:nine-point-stencil}) \cite{GD}. 
\begin{lemma}
   \label{lem:9p} Let the system \eqref{eq:3csyst} be defined on a 2D square lattice. Then, the variable $w$ satisfies the lpBSQ equation  \eqref{eq:dBSQ}. 
\end{lemma}
\prf First  express $y$ using \eqref{eq:3csyst3}  in term of the remaining variable. Second, take respectively the ~$\wt{~}$~ and ~$\wh{~}$~ of the so-obtained $y$, and insert them into \eqref{eq:3csyst3}. This leads to 
\begin{equation}
 \wh{\wt{   (\wh{x}-\wt{x})}} = \frac{(\alpha^3-\beta^3) (\wh{\wh{w}} - 2 \wh{\wt{w}}+ \wt{\wt{w}})  + (\wh{\wh{w}}  - \wh{\wt{w}}) (\wh{\wt{w}} - \wt{\wt{w}}) (\wh{w} \wh{\wh{\wt{w}}}  - \wt{w} \wh{\wt{\wt{w}}}   - w \wh{w}+w\wt{w})}{(\wh{\wh{w}} - \wh{\wt{w}}) (\wh{\wt{w}} -\wt{\wt{w}})}\,.
\end{equation}
Finally, take successively ~$\wt{~}$~ and ~$\wh{~}$~ shifts of \eqref{eq:3csyst1} and replace the term of the ~$\wh{\wt{~}}$~ shift of $\wh{x}-\wt{x}$ from the above formula. This leads to \eqref{eq:dBSQ}.
\finprf
\subsection{Consistency around a $3\times 3 \times 3$ cube}

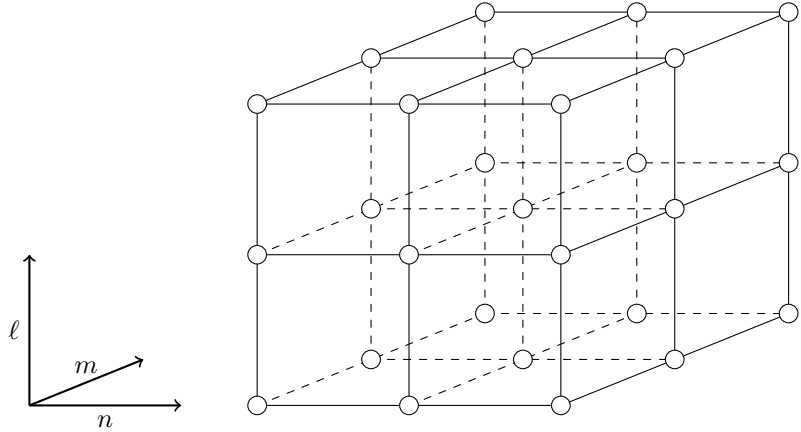
\begin{figure}[htb]   
\centering
		\begin{tikzpicture}[scale=1, decoration={markings,mark=at position 0.55 with {\arrow{latex}}}]
		\tikzstyle{nod1}= [circle, inner sep=0pt, fill=white, minimum size=7pt, draw]
		\tikzstyle{nod}= [circle, inner sep=0pt, fill=black, minimum size=7pt, draw]
			\def\lx{3}%
			\def\ly{1.22}%
			\def\lz{ (sqrt(\x*\x+\y*\y))}%
			\def\l{4}%
			\def\d{2}%

  \draw[thick,->] (-\d/2-\l/2,0) --node[below ]{$n$}  (-\d/2,0)  ;
    \draw[thick,->] (-\d/2-\l/2,0) --node[above ]{$m$}  (-\d/2-\l/2+\lx/2,0+\ly/2)  ;
        \draw[thick,->] (-\d/2-\l/2,0) --node[left ]{$\ell$}  (-\d/2-\l/2,\l/2) ;

			\coordinate (u000) at (0,0); 
			\coordinate (u100) at (\l/2,0); 
			\coordinate (u200) at (2*\l/2,0); 
                \coordinate (u010) at (\lx/2,\ly/2); 
                \coordinate (u110) at (\lx/2+\l/2,\ly/2); 
                \coordinate (u210) at (\lx/2+\l,\ly/2); 
                \coordinate (u020) at (\lx/2+\lx/2,\ly/2+\ly/2); 
                \coordinate (u120) at (\lx/2+\lx/2+\l/2,\ly/2+\ly/2); 
                \coordinate (u220) at (\lx/2+\lx/2+\l,\ly/2+\ly/2); 
	        \coordinate (u001) at (0,\l/2); 
			\coordinate (u101) at (\l/2,\l/2); 
			\coordinate (u201) at (2*\l/2,\l/2); 
                \coordinate (u011) at (\lx/2,\ly/2+\l/2); 
                \coordinate (u111) at (\lx/2+\l/2,\ly/2+\l/2); 
                \coordinate (u211) at (\lx/2+\l,\ly/2+\l/2); 
                \coordinate (u021) at (\lx/2+\lx/2,\ly/2+\ly/2+\l/2); 
                \coordinate (u121) at (\lx/2+\lx/2+\l/2,\ly/2+\ly/2+\l/2); 
                \coordinate (u221) at (\lx/2+\lx/2+\l,\ly/2+\ly/2+\l/2); 
                \coordinate (u002) at (0,\l); 
			\coordinate (u102) at (\l/2,\l); 
			\coordinate (u202) at (2*\l/2,\l); 
                \coordinate (u012) at (\lx/2,\ly/2+\l); 
                \coordinate (u112) at (\lx/2+\l/2,\ly/2+\l); 
                \coordinate (u212) at (\lx/2+\l,\ly/2+\l); 
                \coordinate (u022) at (\lx/2+\lx/2,\ly/2+\ly/2+\l); 
                \coordinate (u122) at (\lx/2+\lx/2+\l/2,\ly/2+\ly/2+\l); 
                \coordinate (u222) at (\lx/2+\lx/2+\l,\ly/2+\ly/2+\l); 

\draw[-] (u000)  --   (u100)--   (u200) ;
\draw[-] (u001)  --   (u101)--   (u201) ;
\draw[-] (u002)  --   (u102)--   (u202) ;


\draw[-] (u000)  --   (u001)--   (u002) ;
\draw[-] (u100)  --   (u101)--   (u102) ;
\draw[-] (u200)  --   (u201)--   (u202) ;

\draw[-] (u002)  --   (u012)--   (u022) ;
\draw[-] (u102)  --   (u112)--   (u122) ;
\draw[-] (u202)  --   (u212)--   (u222) ;
\draw[-] (u012)  --   (u112)--   (u212) ;
\draw[-] (u022)  --   (u122)--   (u222) ;

\draw[-] (u200)  --   (u210)--   (u220) ;
\draw[-] (u201)  --   (u211)--   (u221) ;
\draw[-] (u210)  --   (u211)--   (u212) ;
\draw[-] (u220)  --   (u221)--   (u222) ;
\draw[dashed] (u110)  --   (u111)--   (u112) ;
\draw[dashed] (u010)  --   (u011)--   (u012) ;
\draw[dashed] (u120)  --   (u121)--   (u122) ;
\draw[dashed] (u020)  --   (u021)--   (u022) ;
\draw[dashed] (u010)  --   (u110)--   (u210);
\draw[dashed] (u011)  --   (u111)--   (u211) ;
\draw[dashed] (u021)  --   (u121)--   (u221) ;
\draw[dashed] (u000)  --   (u010)--   (u020) ;
\draw[dashed] (u001)  --   (u011)--   (u021) ;
\draw[dashed] (u101)  --   (u111)--   (u121) ;
\draw[dashed] (u100)  --   (u110)--   (u120) ;
\draw[dashed] (u020)  --   (u120)--   (u220);

                \node[nod1] (u000) at (0,0) 
                {};
                \node[nod1] (u100) at (\l/2,0) 
                {};
                \node[nod1] (u200) at (2*\l/2,0) 
                {};             
                \node[nod1] (u010) at (\lx/2,\ly/2)  {};
                \node[nod1] (u110) at (\lx/2+\l/2,\ly/2)  {};
                \node[nod1] (u210) at (\lx/2+\l,\ly/2)  {};
                \node[nod1] (u020) at (\lx/2+\lx/2,\ly/2+\ly/2) 
                {};
                \node[nod1] (u120) at (\lx/2+\lx/2+\l/2,\ly/2+\ly/2)  {};
                \node[nod1] (u220) at (\lx/2+\lx/2+\l,\ly/2+\ly/2)
                {};
                \node[nod1] (u001) at (0,\l/2)  {};
                \node[nod1] (u101) at (\l/2,\l/2)  {};
                \node[nod1] (u201) at (2*\l/2,\l/2)  {};
                \node[nod1] (u011) at (\lx/2,\ly/2+\l/2)  {};
                \node[nod1] (u111) at (\lx/2+\l/2,\ly/2+\l/2)  {};
                \node[nod1] (u211) at (\lx/2+\l,\ly/2+\l/2)  {};
                \node[nod1] (u021) at (\lx/2+\lx/2,\ly/2+\ly/2+\l/2)   {};
                \node[nod1] (u121) at (\lx/2+\lx/2+\l/2,\ly/2+\ly/2+\l/2)  {};
                \node[nod1] (u221) at (\lx/2+\lx/2+\l,\ly/2+\ly/2+\l/2)   {};
                \node[nod1] (u002) at (0,\l)  {};
                \node[nod1] (u102) at (\l/2,\l)  {};
                \node[nod1] (u202) at (2*\l/2,\l)  {};
                \node[nod1] (u012) at (\lx/2,\ly/2+\l)  {};
                \node[nod1] (u112)  at (\lx/2+\l/2,\ly/2+\l)  {};
                \node[nod1] (u212) at (\lx/2+\l,\ly/2+\l)  {};
                \node[nod1] (u022) at (\lx/2+\lx/2,\ly/2+\ly/2+\l)   {};
                \node[nod1] (u122)  at (\lx/2+\lx/2+\l/2,\ly/2+\ly/2+\l)  {};
                \node[nod1] (u222) at (\lx/2+\lx/2+\l,\ly/2+\ly/2+\l)   {};
		\end{tikzpicture}
		\caption{A $3\times 3\times 3$-vertex cube, the discrete coordinates are fixed. Shifts in independent variables $n,m,\ell \in \mathbb Z$ are respectively denoted by ~$\wt{~}$~, ~$\wh{~}$~, ~$\bar{~}$~, and $\alpha, \beta, \gamma$ are the respective lattice parameters.   } \label{fig:supercube}
	\end{figure}

This section establishes the 3D consistency property of the nine-point equation on a    $3\times 3 \times 3$-vertex cube (see Figure~\ref{fig:supercube}).


Recall $w$ is living on a 3D lattice as in \eqref{eq:wnml}. First, we characterize a set of equations for $w$ obtained from \eqref{eq:3csyst} (or equivalently from \eqref{eq:3csystev}).   
\begin{lemma}\label{lem:35}
Consider the system \eqref{eq:3csyst} in a  three-dimensional lattice, then the discrete field $w$ obeys the following set of equations \begin{subequations}\label{eq:qc}
\begin{align}
\label{eq:QQ}  Q_{\alpha\beta}(w) = 0\,,  \quad Q_{\beta\gamma} (w) =0\,, \quad  Q_{\gamma\alpha}(w) =0\,,\\   \label{eq:CC}C_{\alpha\beta}(w) = 0\,, \quad C_{\beta\gamma} (w) =0\,, \quad  C_{\gamma\alpha}(w) =0\,, 
\end{align}    
\end{subequations}
as well as their shifts in the lattice. Here, 
    \begin{subequations}\label{eq:QCH}
        \begin{align}\label{eq:QCH1}
Q_{\alpha\beta}(w)= &       	\wh{\wh{\wt{\wt{w}}}}(\wh{\wh{\wt{w}}}-\wh{\wt{\wt{w}}})+w(\wh{w}-\wt{w})+(\wt{w}\wh{\wt{\wt{w}}}-\wh{w}\wh{\wh{\wt{w}}})+\frac{\alpha^3-\beta^3}{\wh{\wh{w}}-\wh{\wt{w}}}-\frac{\alpha^3-\beta^3}{\wh{\wt{w}}-\wt{\wt{w}}}\,, \\
\label{eq:QCH2} C_{\alpha\beta}(w)=    &(\wt{\overline{w}} - \wh{\overline{w}})(w-\wh{\wt{\overline{w}}})+\frac{\alpha^3-\gamma^3}{\wt{w}-\overline{w}}-\frac{\beta^3-\gamma^3}{\wh{w}-\overline{w}}\,,
        \end{align}   
and 
            \begin{align}
\label{eq:QC1}            Q_{\beta\gamma} (w) =  Q_{\alpha\beta}(w)\vert_{(\wt{~}, \wh{~}, \bar{~}; \alpha, \beta, \gamma)\to(\wh{~}, \bar{~},\wt{~}; \beta, \gamma, \alpha)}\,,     & \quad Q_{\gamma\alpha} (w) =  Q_{\beta\gamma}(w)\vert_{(\wt{~}, \wh{~}, \bar{~}; \alpha, \beta, \gamma)\to(\wh{~}, \bar{~},\wt{~}; \beta, \gamma, \alpha)}\,, \\   
                     C_{\beta\gamma} (w) =  C_{\alpha\beta}(w)\vert_{(\wt{~}, \wh{~}, \bar{~}; \alpha, \beta, \gamma)\to(\wh{~}, \bar{~},\wt{~}; \beta, \gamma, \alpha)}\,,&         \quad C_{\gamma\alpha} (w) =  C_{\beta\gamma}(w)\vert_{(\wt{~}, \wh{~}, \bar{~}; \alpha, \beta, \gamma)\to(\wh{~}, \bar{~},\wt{~}; \beta, \gamma, \alpha)}\,.
        \end{align}
\end{subequations}
\end{lemma}

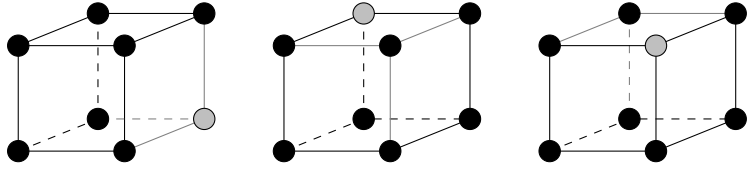
\begin{figure}[htb]  
\centering
		\begin{tikzpicture}[scale=0.7, decoration={markings,mark=at position 0.55 with {\arrow{latex}}}]
		\tikzstyle{nod}= [circle, inner sep=0pt, fill=lightgray, minimum size=8pt, draw]
		\tikzstyle{nod1}= [circle, inner sep=0pt, fill=black, minimum size=8pt, draw]
			\def\lx{3}%
			\def\ly{1.22}%
			\def\lz{ (sqrt(\x*\x+\y*\y))}%
			\def\l{4}%
			\def\ld{1.5}%
            	\def\d{0.8}%
	\coordinate (u00) at (0,0); 
	\coordinate (u10) at (\l/2,0); 
        \coordinate (u01) at (\lx/2,\ly/2); 
        \coordinate (u11) at (\lx/2+\l/2,\ly/2); 
        \coordinate (v00) at (0,0+\l/2); 
	\coordinate (v10) at (\l/2,0+\l/2); 
        \coordinate (v01) at (\lx/2,\ly/2+\l/2); 
        \coordinate (v11) at (\lx/2+\l/2,\ly/2+\l/2); 
\draw[-] (u00)  --   (u10) ;
\draw[dashed] (u01) -- (v01);
\draw[dashed] (u00)  --   (u01) ;
\draw[-] (u00) -- (v00);
\draw[-] (u10) -- (v10);

\draw[-] (v00)  --   (v10) --(v11);
\draw[-] (v00)  --   (v01) --(v11);
\draw[gray] (u10)--(u11);
\draw[gray, dashed] (u11)--(u01);
\draw[gray] (u11)--(v11);
\node[nod1] (u00) at (0,0)  {};
\node[nod1](u10) at (\l/2,0) {};
\node[nod1] (u01) at (\lx/2,\ly/2) {}; 
\node[nod] (u11) at (\lx/2+\l/2,\ly/2) {}; 
\node[nod1] (v00) at (0,0+\l/2)  {};
\node[nod1](v10) at (\l/2,0+\l/2) {};
\node[nod1](v01) at (\lx/2,\ly/2+\l/2){}; 
\node[nod1] (v11) at (\lx/2+\l/2,\ly/2+\l/2) {}; 
	\coordinate (u00) at (\lx/2+\l/2+\ld,0); 
	\coordinate (u10) at (\l/2+\lx/2+\l/2+\ld,0); 
        \coordinate (u01) at (\lx/2+\lx/2+\l/2+\ld,\ly/2); 
        \coordinate (u11) at (\lx/2+\l/2+\lx/2+\l/2+\ld,\ly/2); 
        \coordinate (v00) at (0+\lx/2+\l/2+\ld,0+\l/2); 
	\coordinate (v10) at (\l/2+\lx/2+\l/2+\ld,0+\l/2); 
        \coordinate (v01) at (\lx/2+\lx/2+\l/2+\ld,\ly/2+\l/2); 
        \coordinate (v11) at (\lx/2+\l/2+\lx/2+\l/2+\ld,\ly/2+\l/2); %
\draw[-] (u00)  --   (u10) -- (u11) --(v11) -- (v01) --(v00) --(u00) ;
\draw[dashed] (u00) -- (u01) -- (u11);
\draw[dashed] (u01) -- (v01);
\draw[gray] (v10)--(u10);
\draw[gray] (v10)--(v11);
\draw[gray] (v10)--(v00);
\node[nod1] (u00) at (0+\lx/2+\l/2+\ld,0)  {};
\node[nod1](u10) at (\l/2+\lx/2+\l/2+\ld,0) {};
\node[nod1] (u01) at (\lx/2+\lx/2+\l/2+\ld,\ly/2) {}; 
\node[nod1] (u11) at (\lx/2+\l/2+\lx/2+\l/2+\ld,\ly/2) {}; 
\node[nod1] (v00) at (0+\lx/2+\l/2+\ld,0+\l/2)  {};
\node[nod1](v10) at (\l/2+\lx/2+\l/2+\ld,0+\l/2) {};
\node[nod](v01) at (\lx/2+\lx/2+\l/2+\ld,\ly/2+\l/2){}; 
\node[nod1] (v11) at (\lx/2+\l/2+\lx/2+\l/2+\ld,\ly/2+\l/2) {}; 
	\coordinate (u00) at (0+\lx/2+\l/2+\lx/2+\l/2+\ld+\ld,0); 
	\coordinate (u10) at (\l/2+\lx/2+\l/2+\lx/2+\l/2+\ld+\ld,0); 
        \coordinate (u01) at (\lx/2+\lx/2+\l/2+\lx/2+\l/2+\ld+\ld,\ly/2); 
        \coordinate (u11) at (\lx/2+\l/2+\lx/2+\l/2+\lx/2+\l/2+\ld+\ld,\ly/2); 
        \coordinate (v00) at (0+\lx/2+\l/2+\lx/2+\l/2+\ld+\ld,0+\l/2); 
	\coordinate (v10) at (\l/2+\lx/2+\l/2+\lx/2+\l/2+\ld+\ld,0+\l/2); 
        \coordinate (v01) at (\lx/2+\lx/2+\l/2+\lx/2+\l/2+\ld+\ld,\ly/2+\l/2); 
        \coordinate (v11) at (\lx/2+\l/2+\lx/2+\l/2+\lx/2+\l/2+\ld+\ld,\ly/2+\l/2); 

\draw[gray] (v01)--(v11);
\draw[gray, dashed] (v01)--(u01);
\draw[gray] (v01)--(v00);
\draw[-] (u00)--(v00) --(v10)--(v11)--(u11)--(u10) --(u00);
\draw[-] (u10)--(v10);
\draw[dashed] (u00) -- (u01) -- (u11) ; 
\node[nod1] (u00) at (0+\lx/2+\l/2+\lx/2+\l/2+\ld+\ld,0)  {};
\node[nod1](u10) at (\l/2+\lx/2+\l/2+\lx/2+\l/2+\ld+\ld,0) {};
\node[nod1] (u01) at (\lx/2+\lx/2+\l/2+\lx/2+\l/2+\ld+\ld,\ly/2) {}; 
\node[nod1] (u11) at (\lx/2+\l/2+\lx/2+\l/2+\lx/2+\l/2+\ld+\ld,\ly/2) {}; 
\node[nod1] (v00) at (0+\lx/2+\l/2+\lx/2+\l/2+\ld+\ld,0+\l/2)  {};
\node[nod](v10) at (\l/2+\lx/2+\l/2+\lx/2+\l/2+\ld+\ld,0+\l/2) {};
\node[nod1](v01) at (\lx/2+\lx/2+\l/2+\lx/2+\l/2+\ld+\ld,\ly/2+\l/2){}; 
\node[nod1] (v11) at (\lx/2+\l/2+\lx/2+\l/2+\lx/2+\l/2+\ld+\ld,\ly/2+\l/2) {};

\end{tikzpicture}
		\caption{The supporting lattices of the equations $C_{\alpha\beta}(w)=0$, $C_{\beta\gamma}(w)=0$ and $C_{\gamma\alpha}(w)=0$ } are respectively depicted from left to right with the {\em gray dot} representing the removed vertex.  \label{fig:eqs}
\end{figure}

Let us comment on the derivations as well as the elementary supporting lattices of the above equations for $w$. 
\begin{itemize}
    \item The first equation  $ {Q}_{\alpha\beta} =0$  is the nine point equation \eqref{eq:dBSQ} defined on a $3\times 3$-vertex (nine-point) square lattice in the $(n,m)$ plane. Its derivation was given in Lemma \ref{lem:35}. In a $3\times 3\times 3$ cube (see  Figure~\ref{fig:supercube}), there are extra two equations in the $(n,m)$ plane shifted in the $\ell$ direction, \ie $ \overline{Q}_{\alpha\beta} =0$, $ \overline{\overline{Q}}_{\alpha\beta} = 0$. One also has $Q_{\beta\gamma} (w) = 0$, $Q_{\gamma\alpha}= 0$ respectively on the $(m,\ell)$ and $(\ell,\alpha)$ planes and their shifted versions: $\wt{Q}_{\beta\gamma} (w) = 0$, $\wt{\wt{Q}}_{\beta\gamma} (w) = 0$, $\wh{Q}_{\gamma\alpha}= 0$, $\wh{\wh{Q}}_{\gamma\alpha}= 0$. There are in total $9$ nine-point equations on a  $3\times 3\times 3$  cube. 
    \item We called the second equation $ C_{\alpha\beta} (w)  =0$ and its variants  $ C_{\beta\gamma} (w)  =0$ and $ C_{\gamma\alpha} (w)  =0$ {\em 3D companions of lpBSQ}.  They  are defined on an elementary cube with one vertex removed (see Figure~\ref{fig:eqs}). It can be derived as follows: consider \eqref{eq:3csyst3} in the $(m,\ell)$ and $(\ell, n)$ planes  
    \begin{equation}
        \widehat{\overline{x}} =w  \widehat{\overline{w}} -y + \frac{\gamma^3-\beta^3}{\overline{w} -\wh{w}}\,,\quad         \wt{\overline{x}} =w  \wt{\overline{w}} -y + \frac{\alpha^3-\gamma^3}{\wt{w} -\overline{w}}\,. 
    \end{equation}
On the one hand, their difference yields
     \begin{equation}
    \wt{\overline{x}} -       \widehat{\overline{x}} =w  \wt{\overline{w}}  - w  \widehat{\overline{w}} +\frac{\alpha^3-\gamma^3}{\wt{w} -\overline{w}}
    -  \frac{\gamma^3-\beta^3}{\overline{w} -\wh{w}}\,, \end{equation}
on the other  taking the ~$\bar{~}$~ shift of \eqref{eq:3csyst1},  
 one gets 
\begin{equation}
    \wh{\overline{x}}-\wt{\overline{x}}=     (\wh{\overline{w}}-\wt{\overline{w}})\wh{\wt{\overline{w}}}\,. 
\end{equation}
Combining the above two equations leads to \eqref{eq:QCH2}.

\end{itemize}

There are two comments concerning the nature of the 3D companions of lpBSQ $C_{\alpha\beta} (w)=0$ and its variants $C_{\beta\gamma} (w)=0$, $ C_{\alpha\beta} (w) = 0$. 

\begin{rmk}
One can show $C_{\alpha\beta} (w)+C_{\beta\gamma} (w) + C_{\alpha\beta} (w) = 0$. This implies that among the three equations  $C_{\alpha\beta} (w) = 0$, $C_{\beta\gamma} (w)=0$ and $ C_{\alpha\beta} (w) = 0$, one is a consequence of the other two. Moreover, they are consistent with
the lattice (KP) equation \cite{HM, ABS_KP} \begin{equation}\label{eq:hw}
 H(w)= \widehat{\widetilde{w}}(\widehat{w}-\widetilde{w})+\widehat{\overline{w}}(\overline{w}-\widehat{w})+   	\widetilde{\overline{w}}(\widetilde{w}-\overline{w}) =0\,,
\end{equation}
which is implicitly present in the whole discrete Gel’fand Dikii hierarchy \cite{GD}. 
\label{rm:36}
\end{rmk}
\begin{rmk}
The lpBSQ variable $w$ also obeys {\em L-shaped} equations defined on a 3D nine-point lattice as depicted in Figure~\ref{fig:L-shape}. This is obtained by combining 3D companions of lpBSQ  \eqref{eq:CC} and the lattice KP equations \eqref{eq:hw}. For instance, take $C_{\alpha\beta}=0$ and $\overline{H}(w)=0$ which is the ~$\bar{~}$~ shift of \eqref{eq:hw}:
\begin{subequations}
\begin{align}
C_{\alpha\beta}(w) =&    (\wt{\overline{w}} - \wh{\overline{w}})(w-\wh{\wt{\overline{w}}})+\frac{\alpha^3-\gamma^3}{\wt{w}-\overline{w}}-\frac{\beta^3-\gamma^3}{\wh{w}-\overline{w}} = 0\,, \\
  \overline{H}(w) =& \widehat{\widetilde{\overline{w}}}(\widehat{\overline{w}}-\widetilde{\overline{w}})+\widehat{\overline{\overline{w}}}(\overline{\overline{w}}-\widehat{\overline{w}})+   	\widetilde{\overline{\overline{w}}}(\widetilde{\overline{w}}-\overline{\overline{w}}) = 0\,.
\end{align}    
\end{subequations}
by eliminating $\wt{\wh{\overline{w}}}$ in the above equations, one gets
\begin{equation}
  \cL_{\alpha\beta}(w)=  (\wt{\overline{w}}-\wh{\overline{w}})w+(\wt{\overline{\overline{w}}}-\wh{\overline{\overline{w}}})\overline{\overline{w}}+\wh{\overline{w}}\,\wh{\overline{\overline{w}}}-\wt{\overline{w}}\,\wt{\overline{\overline{w}}}+\frac{\alpha^3-\gamma^3}{\wt{w}-\overline{w}}-\frac{\beta^3-\gamma^3}{\wh{w}-\overline{w}}=0\,, 
\end{equation}
which is defined on a L-shaped 3D nine-point lattice involving $(\wt{w},w, \wh{w})$,  $(\wt{\overline{w}},\overline{w}, \wh{\overline{w}})$ and $(\wt{\overline{\overline{w}}},\overline{\overline{w}}, \wh{\overline{\overline{w}}})$. Similarly, one can obtain L-shaped equations in other orientations as
\begin{equation}
    \cL_{\beta\gamma}(w) = \cL_{\alpha\beta}(w)\vert_{(\wt{~}, \wh{~}, \bar{~}; \alpha, \beta, \gamma)\to(\wh{~}, \bar{~},\wt{~}; \beta, \gamma, \alpha)}\,,\quad     \cL_{\gamma\alpha}(w) = \cL_{\beta\gamma}(w)\vert_{(\wt{~}, \wh{~}, \bar{~}; \alpha, \beta, \gamma)\to(\wh{~}, \bar{~},\wt{~}; \beta, \gamma, \alpha)}\,.
\end{equation}
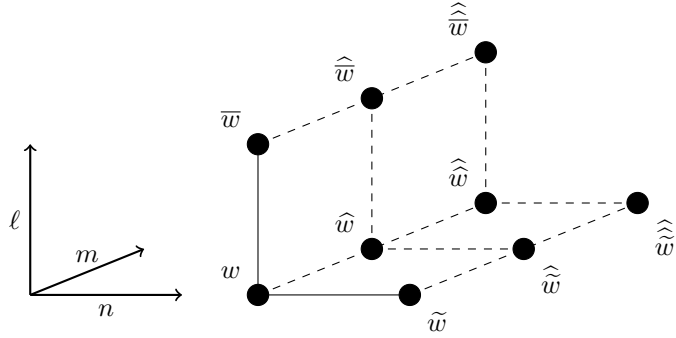
\begin{figure}[htb]  
\centering
		\begin{tikzpicture}[scale=1, decoration={markings,mark=at position 0.55 with {\arrow{latex}}}]
   \tikzstyle{nod}= [circle, inner sep=0pt, fill=white, minimum size=8pt, draw]
\tikzstyle{nod1}= [circle, inner sep=0pt, fill=black, minimum size=8pt, draw]
\tikzstyle{nod2}= [circle, inner sep=0pt, fill=red, minimum size=8pt, draw]
\tikzstyle{nod3}= [circle, inner sep=0pt, fill=yellow, minimum size=8pt, draw]
\tikzstyle{nodsb}= [rectangle, inner sep=0pt, fill=white, minimum size=8pt, draw]
\tikzstyle{nodsw}= [rectangle, inner sep=0pt, fill=white, minimum size=8pt, draw]
\tikzstyle{nodsb}= [rectangle, inner sep=0pt, fill=black, minimum size=8pt, draw]
\tikzstyle{nodt}= [isosceles triangle, isosceles triangle apex angle=60, shape border rotate=90, draw,  fill=white, minimum
width=8pt,minimum height=8pt, inner sep=0pt, draw]
\tikzstyle{nodtb}= [isosceles triangle, isosceles triangle apex angle=60, shape border rotate=90, draw,  fill=black, minimum
width=8pt,minimum height=8pt, inner sep=0pt, draw]
\tikzstyle{nodrt}= [isosceles triangle, isosceles triangle apex angle=60, shape border rotate=270, draw,  fill=white, minimum
width=8pt,minimum height=8pt, inner sep=0pt, draw]
\tikzstyle{nodd}= [diamond, fill=white, isosceles triangle stretches,shape border rotate=270,minimum size=6pt, draw]

			\def\lx{3}%
			\def\ly{1.22}%
			\def\lz{ (sqrt(\x*\x+\y*\y))}%
			\def\l{4}%
			\def\d{2}%
  \draw[thick,->] (-\d/2-\l/2,0) --node[below ]{$n$}  (-\d/2,0)  ;
    \draw[thick,->] (-\d/2-\l/2,0) --node[above ]{$m$}  (-\d/2-\l/2+\lx/2,0+\ly/2)  ;
        \draw[thick,->] (-\d/2-\l/2,0) --node[left ]{$\ell$}  (-\d/2-\l/2,\l/2) ;
            
			\coordinate (u000) at (0,0); 
			\coordinate (u100) at (\l/2,0); 
			\coordinate (u200) at (2*\l/2,0); 
                \coordinate (u010) at (\lx/2,\ly/2); 
                \coordinate (u110) at (\lx/2+\l/2,\ly/2); 
                \coordinate (u210) at (\lx/2+\l,\ly/2); 
                \coordinate (u020) at (\lx/2+\lx/2,\ly/2+\ly/2); 
                \coordinate (u120) at (\lx/2+\lx/2+\l/2,\ly/2+\ly/2); 
                \coordinate (u220) at (\lx/2+\lx/2+\l,\ly/2+\ly/2); 
	        \coordinate (u001) at (0,\l/2); 
			\coordinate (u101) at (\l/2,\l/2); 
			\coordinate (u201) at (2*\l/2,\l/2); 
                \coordinate (u011) at (\lx/2,\ly/2+\l/2); 
                \coordinate (u111) at (\lx/2+\l/2,\ly/2+\l/2); 
                \coordinate (u211) at (\lx/2+\l,\ly/2+\l/2); 
                \coordinate (u021) at (\lx/2+\lx/2,\ly/2+\ly/2+\l/2); 
                \coordinate (u121) at (\lx/2+\lx/2+\l/2,\ly/2+\ly/2+\l/2); 
                \coordinate (u221) at (\lx/2+\lx/2+\l,\ly/2+\ly/2+\l/2); 
                \coordinate (u002) at (0,\l); 
			\coordinate (u102) at (\l/2,\l); 
			\coordinate (u202) at (2*\l/2,\l); 
                \coordinate (u012) at (\lx/2,\ly/2+\l); 
                \coordinate (u112) at (\lx/2+\l/2,\ly/2+\l); 
                \coordinate (u212) at (\lx/2+\l,\ly/2+\l); 
                \coordinate (u022) at (\lx/2+\lx/2,\ly/2+\ly/2+\l); 
                \coordinate (u122) at (\lx/2+\lx/2+\l/2,\ly/2+\ly/2+\l); 
                \coordinate (u222) at (\lx/2+\lx/2+\l,\ly/2+\ly/2+\l); 

\draw[-] (u000)  --   (u100)  ;
\draw[-] (u000)  --   (u001) ;
\draw[dashed] (u010)  --   (u011)  ;
\draw[dashed] (u020)  --   (u021) ;
\draw[dashed] (u010)  --   (u110) ;
\draw[dashed] (u000)  --   (u010)--   (u020) ;
\draw[dashed] (u001)  --   (u011)--   (u021) ;
\draw[dashed] (u100)  --   (u110)--   (u120) ;
\draw[dashed] (u020)  --   (u120) ;

                \node[nod1] (u000) at (0,0) [label=above left: ${w}$] 
                {};
                \node[nod1] (u100) at (\l/2,0) [label=below right: $\widetilde{w}$] 
                {};
                \node[nod1] (u010) at (\lx/2,\ly/2) [label=above left: $\wh{w}$]  {};
                \node[nod1] (u110) at (\lx/2+\l/2,\ly/2)  [label=below  right: $\wh{\wt{w}}$]   {};
                \node[nod1] (u020) at (\lx/2+\lx/2,\ly/2+\ly/2)  [label=above left: $\widehat{\widehat{w}}$] 
                {};
                \node[nod1] (u120) at (\lx/2+\lx/2+\l/2,\ly/2+\ly/2)  [label=below right: $\widehat{\widehat{\wt{w}}}$] {};
                {};
                \node[nod1] (u001) at (0,\l/2)  [label=above left: $\overline{w}$] {};
                \node[nod1] (u011) at (\lx/2,\ly/2+\l/2) [label=above left: $\wh{\overline{w}}$]  {};
                     \node[nod1] (u021) at (\lx/2+\lx/2,\ly/2+\ly/2+\l/2)   [label=above  left: $\wh{\wh{\overline{w}}}$]  {};
		\end{tikzpicture}
		\caption{The L-shaped BSQ equation $\cL_{\gamma\alpha} (w)= 0$ defined on a 3D nine-point lattice in the $3\times 3\times 3$ cube which can be obtained by eliminating $\wt{\wh{\overline{w}}}$ in $C_{\gamma\alpha}(w)=0$ and $\wh{H}(w) = 0$. } \label{fig:L-shape}
	\end{figure}
\end{rmk}

Let us describe the consistency property of the equations \eqref{eq:QCH} given in Lemma \ref{lem:35} around a $3\times 3\times 3$ cube. In order to do so, one needs to characterize two type of evolutionary equations: {\em planar equation} and {\em cubic system}.  
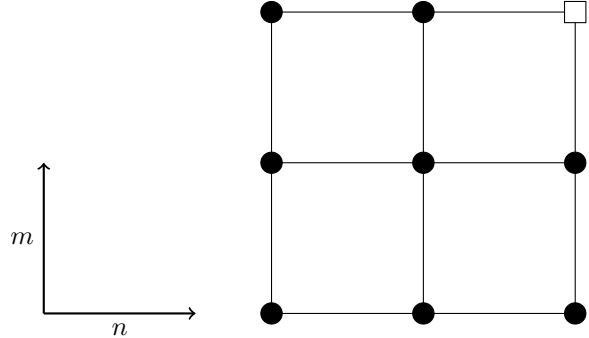
\begin{figure}[htb] 
    \centering
  
\centering
		\begin{tikzpicture}[scale=1, decoration={markings,mark=at position 0.55 with {\arrow{latex}}}]
		\tikzstyle{nod}= [circle, inner sep=0pt, fill=white, minimum size=8pt, draw]
		\tikzstyle{nod1}= [circle, inner sep=0pt, fill=black, minimum size=8pt, draw]
            \tikzstyle{nod2}= [circle, inner sep=0pt, fill=red, minimum size=8pt, draw]
            \tikzstyle{nod3}= [circle, inner sep=0pt, fill=yellow, minimum size=8pt, draw]
  \tikzstyle{nodsb}= [rectangle, inner sep=0pt, fill=white, minimum size=8pt, draw]
            \tikzstyle{nodsw}= [rectangle, inner sep=0pt, fill=white, minimum size=8pt, draw]
              \tikzstyle{nodsb}= [rectangle, inner sep=0pt, fill=gray, minimum size=8pt, draw]
            \tikzstyle{nodt}= [isosceles triangle, isosceles triangle apex angle=60, shape border rotate=90, draw,  fill=white, minimum
width=8pt,minimum height=8pt, inner sep=0pt, draw]
 \tikzstyle{nodtb}= [isosceles triangle, isosceles triangle apex angle=60, shape border rotate=90, draw,  fill=gray, minimum
width=8pt,minimum height=8pt, inner sep=0pt, draw]
 \tikzstyle{nodrt}= [isosceles triangle, isosceles triangle apex angle=60, shape border rotate=270, draw,  fill=white, minimum
width=8pt,minimum height=8pt, inner sep=0pt, draw]
           \tikzstyle{nodd}= [diamond, fill=white, isosceles triangle stretches,shape border rotate=270,minimum size=6pt, draw]
             \tikzstyle{noddb}= [diamond, fill=gray, isosceles triangle stretches,shape border rotate=270,minimum size=6pt, draw]
			\def\lx{3}%
			\def\ly{1.22}%
			\def\lz{ (sqrt(\x*\x+\y*\y))}%
			\def\l{4}%
			\def\d{2}%
  \draw[thick,->] (-\d/2-\l/2,0) --node[below ]{$n$}  (-\d/2,0)  ;

        \draw[thick,->] (-\d/2-\l/2,0) --node[left ]{$m$}  (-\d/2-\l/2,\l/2) ;
            
			\coordinate (u000) at (0,0); 
			\coordinate (u100) at (\l/2,0); 
			\coordinate (u200) at (2*\l/2,0); 
                \coordinate (u010) at (\lx/2,\ly/2); 
                \coordinate (u110) at (\lx/2+\l/2,\ly/2); 
                \coordinate (u210) at (\lx/2+\l,\ly/2); 
                \coordinate (u020) at (\lx/2+\lx/2,\ly/2+\ly/2); 
                \coordinate (u120) at (\lx/2+\lx/2+\l/2,\ly/2+\ly/2); 
                \coordinate (u220) at (\lx/2+\lx/2+\l,\ly/2+\ly/2); 
	        \coordinate (u001) at (0,\l/2); 
			\coordinate (u101) at (\l/2,\l/2); 
			\coordinate (u201) at (2*\l/2,\l/2); 
                \coordinate (u011) at (\lx/2,\ly/2+\l/2); 
                \coordinate (u111) at (\lx/2+\l/2,\ly/2+\l/2); 
                \coordinate (u211) at (\lx/2+\l,\ly/2+\l/2); 
                \coordinate (u021) at (\lx/2+\lx/2,\ly/2+\ly/2+\l/2); 
                \coordinate (u121) at (\lx/2+\lx/2+\l/2,\ly/2+\ly/2+\l/2); 
                \coordinate (u221) at (\lx/2+\lx/2+\l,\ly/2+\ly/2+\l/2); 
                \coordinate (u002) at (0,\l); 
			\coordinate (u102) at (\l/2,\l); 
			\coordinate (u202) at (2*\l/2,\l); 
                \coordinate (u012) at (\lx/2,\ly/2+\l); 
                \coordinate (u112) at (\lx/2+\l/2,\ly/2+\l); 
                \coordinate (u212) at (\lx/2+\l,\ly/2+\l); 
                \coordinate (u022) at (\lx/2+\lx/2,\ly/2+\ly/2+\l); 
                \coordinate (u122) at (\lx/2+\lx/2+\l/2,\ly/2+\ly/2+\l); 
                \coordinate (u222) at (\lx/2+\lx/2+\l,\ly/2+\ly/2+\l); 

\draw[-] (u000)  --   (u100)--   (u200) ;
\draw[-] (u001)  --   (u101)--   (u201) ;
\draw[-] (u002)  --   (u102)--(u202) ;


\draw[-] (u000)  --   (u001)--   (u002) ;
\draw[-] (u100)  --   (u101)--   (u102) ;
\draw[-] (u200)  --   (u201)--   (u202) ;

\node[nod1] (u000) at (0,0) {};
\node[nod1] (u100) at (\l/2,0){};
\node[nod1] (u200) at (2*\l/2,0){};
\node[nod1] (u001) at (0,\l/2)  {};
\node[nod1] (u101) at (\l/2,\l/2)  {};
\node[nod1] (u201) at (2*\l/2,\l/2)  {};
\node[nod1] (u002) at (0,\l)  {};
\node[nod1] (u102) at (\l/2,\l)  {};
\node[nodsw] (u202) at (2*\l/2,\l)  {};

\end{tikzpicture}
	\caption{Planar equation: consider a nine-point equation $Q_{\alpha\beta}(w) =0$ in the $(n,m)$ plane. With generic initial values on the $8$ {\em black dots} as well as the lattice parameters, the value on the {\em white square} can be uniquely obtained.}  \label{fig:nine-point}
\end{figure}
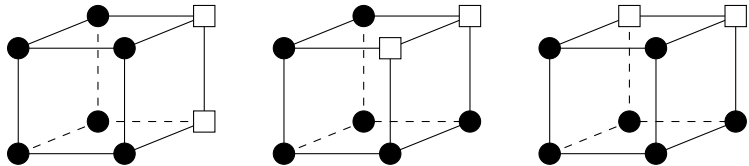
\begin{figure}[htb]  
\centering
		\begin{tikzpicture}[scale=0.7, decoration={markings,mark=at position 0.55 with {\arrow{latex}}}]
		\tikzstyle{nod}= [circle, inner sep=0pt, fill=lightgray, minimum size=8pt, draw]
		\tikzstyle{nod1}= [circle, inner sep=0pt, fill=black, minimum size=8pt, draw]
         \tikzstyle{nodsw}= [rectangle, inner sep=0pt, fill=white, minimum size=8pt, draw]
			\def\lx{3}%
			\def\ly{1.22}%
			\def\lz{ (sqrt(\x*\x+\y*\y))}%
			\def\l{4}%
			\def\ld{1.5}%
            	\def\d{0.8}%
	\coordinate (u00) at (0,0); 
	\coordinate (u10) at (\l/2,0); 
        \coordinate (u01) at (\lx/2,\ly/2); 
        \coordinate (u11) at (\lx/2+\l/2,\ly/2); 
        \coordinate (v00) at (0,0+\l/2); 
	\coordinate (v10) at (\l/2,0+\l/2); 
        \coordinate (v01) at (\lx/2,\ly/2+\l/2); 
        \coordinate (v11) at (\lx/2+\l/2,\ly/2+\l/2); 
\draw[-] (u00)  --   (u10) ;
\draw[dashed] (u01) -- (v01);
\draw[dashed] (u00)  --   (u01) ;
\draw[-] (u00) -- (v00);
\draw[-] (u10) -- (v10);

\draw[-] (v00)  --   (v10) --(v11);
\draw[-] (v00)  --   (v01) --(v11);
\draw[-] (u10)--(u11);
\draw[dashed] (u11)--(u01);
\draw[-] (u11)--(v11);
\node[nod1] (u00) at (0,0)  {};
\node[nod1](u10) at (\l/2,0) {};
\node[nod1] (u01) at (\lx/2,\ly/2) {}; 
\node[nodsw] (u11) at (\lx/2+\l/2,\ly/2) {}; 
\node[nod1] (v00) at (0,0+\l/2)  {};
\node[nod1](v10) at (\l/2,0+\l/2) {};
\node[nod1](v01) at (\lx/2,\ly/2+\l/2){}; 
\node[nodsw] (v11) at (\lx/2+\l/2,\ly/2+\l/2) {}; 
	\coordinate (u00) at (\lx/2+\l/2+\ld,0); 
	\coordinate (u10) at (\l/2+\lx/2+\l/2+\ld,0); 
        \coordinate (u01) at (\lx/2+\lx/2+\l/2+\ld,\ly/2); 
        \coordinate (u11) at (\lx/2+\l/2+\lx/2+\l/2+\ld,\ly/2); 
        \coordinate (v00) at (0+\lx/2+\l/2+\ld,0+\l/2); 
	\coordinate (v10) at (\l/2+\lx/2+\l/2+\ld,0+\l/2); 
        \coordinate (v01) at (\lx/2+\lx/2+\l/2+\ld,\ly/2+\l/2); 
        \coordinate (v11) at (\lx/2+\l/2+\lx/2+\l/2+\ld,\ly/2+\l/2); %
\draw[-] (u00)  --   (u10) -- (u11) --(v11) -- (v01) --(v00) --(u00) ;
\draw[dashed] (u00) -- (u01) -- (u11);
\draw[dashed] (u01) -- (v01);
\draw[-] (v10)--(u10);
\draw[-] (v10)--(v11);
\draw[-] (v10)--(v00);
\node[nod1] (u00) at (0+\lx/2+\l/2+\ld,0)  {};
\node[nod1](u10) at (\l/2+\lx/2+\l/2+\ld,0) {};
\node[nod1] (u01) at (\lx/2+\lx/2+\l/2+\ld,\ly/2) {}; 
\node[nod1] (u11) at (\lx/2+\l/2+\lx/2+\l/2+\ld,\ly/2) {}; 
\node[nod1] (v00) at (0+\lx/2+\l/2+\ld,0+\l/2)  {};
\node[nodsw](v10) at (\l/2+\lx/2+\l/2+\ld,0+\l/2) {};
\node[nod1](v01) at (\lx/2+\lx/2+\l/2+\ld,\ly/2+\l/2){}; 
\node[nodsw] (v11) at (\lx/2+\l/2+\lx/2+\l/2+\ld,\ly/2+\l/2) {}; 
	\coordinate (u00) at (0+\lx/2+\l/2+\lx/2+\l/2+\ld+\ld,0); 
	\coordinate (u10) at (\l/2+\lx/2+\l/2+\lx/2+\l/2+\ld+\ld,0); 
        \coordinate (u01) at (\lx/2+\lx/2+\l/2+\lx/2+\l/2+\ld+\ld,\ly/2); 
        \coordinate (u11) at (\lx/2+\l/2+\lx/2+\l/2+\lx/2+\l/2+\ld+\ld,\ly/2); 
        \coordinate (v00) at (0+\lx/2+\l/2+\lx/2+\l/2+\ld+\ld,0+\l/2); 
	\coordinate (v10) at (\l/2+\lx/2+\l/2+\lx/2+\l/2+\ld+\ld,0+\l/2); 
        \coordinate (v01) at (\lx/2+\lx/2+\l/2+\lx/2+\l/2+\ld+\ld,\ly/2+\l/2); 
        \coordinate (v11) at (\lx/2+\l/2+\lx/2+\l/2+\lx/2+\l/2+\ld+\ld,\ly/2+\l/2); 

\draw[-] (v01)--(v11);
\draw[dashed] (v01)--(u01);
\draw[-] (v01)--(v00);
\draw[-] (u00)--(v00) --(v10)--(v11)--(u11)--(u10) --(u00);
\draw[-] (u10)--(v10);
\draw[dashed] (u00) -- (u01) -- (u11) ; 
\node[nod1] (u00) at (0+\lx/2+\l/2+\lx/2+\l/2+\ld+\ld,0)  {};
\node[nod1](u10) at (\l/2+\lx/2+\l/2+\lx/2+\l/2+\ld+\ld,0) {};
\node[nod1] (u01) at (\lx/2+\lx/2+\l/2+\lx/2+\l/2+\ld+\ld,\ly/2) {}; 
\node[nod1] (u11) at (\lx/2+\l/2+\lx/2+\l/2+\lx/2+\l/2+\ld+\ld,\ly/2) {}; 
\node[nod1] (v00) at (0+\lx/2+\l/2+\lx/2+\l/2+\ld+\ld,0+\l/2)  {};
\node[nod1](v10) at (\l/2+\lx/2+\l/2+\lx/2+\l/2+\ld+\ld,0+\l/2) {};
\node[nodsw](v01) at (\lx/2+\lx/2+\l/2+\lx/2+\l/2+\ld+\ld,\ly/2+\l/2){}; 
\node[nodsw] (v11) at (\lx/2+\l/2+\lx/2+\l/2+\lx/2+\l/2+\ld+\ld,\ly/2+\l/2) {}; 
		\end{tikzpicture}
		\caption{Cubic system: having generic initial data on the {\em black dots} as well as the lattice parameters, one can uniquely obtain the values on the {\em white squares}.} \label{fig:cacBSQ}
	\end{figure}
\begin{description}
    \item[Planar equation: evolutionary system on a nine-point square lattice.] Given a nine-point equation, the initial-value problem on a nine-point square lattice requires known values on $8$ vertices as well as the lattice parameters. This is explained in Figure~\ref{fig:nine-point}.  
    
    \item[Cubic system: evolutionary system on an elementary cube.]  Given  $C_{\alpha\beta} (w) = 0$, $C_{\beta\gamma} (w)=0$ and $ C_{\alpha\beta} (w) = 0$ on an elementary cube, there are three ways to set up initial-value problems as depicted in Figure~\ref{fig:cacBSQ}. Precisely, taking $C_{\alpha\beta} (w) = 0$ and $C_{\beta\gamma} (w)=0$:
    \begin{equation}
        (\wt{\overline{w}} - \wh{\overline{w}})(w-\wh{\wt{\overline{w}}})+\frac{\alpha^3-\gamma^3}{\wt{w}-\overline{w}}-\frac{\beta^3-\gamma^3}{\wh{w}-\overline{w}} = 0\,,~~(\wh{\wt{w}} - \wt{\overline{w}})(w-\wh{\wt{\overline{w}}})+\frac{\beta^3-\alpha^3}{\wh{w}-\wt{w}}-\frac{\gamma^3-\alpha^3}{\overline{w}-\wt{w}}=0\,,
    \end{equation}
one gets three systems (depicted respectively from left to right in Figure~\ref{fig:cacBSQ}):\begin{enumerate}
    \item expressing uniquely $(\wh{\wt{w}}, \wh{\wt{\overline{w}}}) $ in terms of $(w, \wt{w}, \wh{w}, \overline{w}, \wt{\overline{w}}, \wh{\overline{w}})$ as well as the lattice parameters;
    \item expressing uniquely $(\wt{\overline{w}}, \wh{\wt{\overline{w}}}) $ in terms of $(w, \wt{w}, \wh{w}, \overline{w}, \wh{\overline{w}},\wt{\overline{w}} )$ as well as the lattice parameters;
    \item expressing uniquely $(\wh{\overline{w}}, \wh{\wt{\overline{w}}}) $ in terms of $(w, \wt{w}, \wh{w}, \overline{w},\wt{\overline{w}}, \wh{\wt{w}} )$ as well as the lattice parameters.
\end{enumerate}
\end{description}
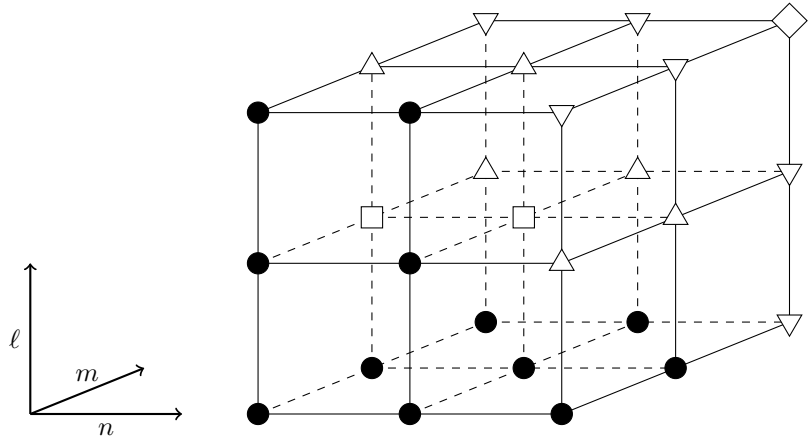
\begin{figure}[htb]  
\centering
		\begin{tikzpicture}[scale=1, decoration={markings,mark=at position 0.55 with {\arrow{latex}}}]
		\tikzstyle{nod}= [circle, inner sep=0pt, fill=white, minimum size=8pt, draw]
		\tikzstyle{nod1}= [circle, inner sep=0pt, fill=black, minimum size=8pt, draw]
            \tikzstyle{nod2}= [circle, inner sep=0pt, fill=red, minimum size=8pt, draw]
            \tikzstyle{nod3}= [circle, inner sep=0pt, fill=yellow, minimum size=8pt, draw]
  \tikzstyle{nodsb}= [rectangle, inner sep=0pt, fill=white, minimum size=8pt, draw]
            \tikzstyle{nodsw}= [rectangle, inner sep=0pt, fill=white, minimum size=8pt, draw]
              \tikzstyle{nodsb}= [rectangle, inner sep=0pt, fill=gray, minimum size=8pt, draw]
            \tikzstyle{nodt}= [isosceles triangle, isosceles triangle apex angle=60, shape border rotate=90, draw,  fill=white, minimum
width=8pt,minimum height=8pt, inner sep=0pt, draw]
 \tikzstyle{nodtb}= [isosceles triangle, isosceles triangle apex angle=60, shape border rotate=90, draw,  fill=gray, minimum
width=8pt,minimum height=8pt, inner sep=0pt, draw]
 \tikzstyle{nodrt}= [isosceles triangle, isosceles triangle apex angle=60, shape border rotate=270, draw,  fill=white, minimum
width=8pt,minimum height=8pt, inner sep=0pt, draw]
           \tikzstyle{nodd}= [diamond, fill=white, isosceles triangle stretches,shape border rotate=270,minimum size=6pt, draw]
             \tikzstyle{noddb}= [diamond, fill=white, isosceles triangle stretches,shape border rotate=270,minimum size=6pt, draw]
			\def\lx{3}%
			\def\ly{1.22}%
			\def\lz{ (sqrt(\x*\x+\y*\y))}%
			\def\l{4}%
			\def\d{2}%
  \draw[thick,->] (-\d/2-\l/2,0) --node[below ]{$n$}  (-\d/2,0)  ;
    \draw[thick,->] (-\d/2-\l/2,0) --node[above ]{$m$}  (-\d/2-\l/2+\lx/2,0+\ly/2)  ;
        \draw[thick,->] (-\d/2-\l/2,0) --node[left ]{$\ell$}  (-\d/2-\l/2,\l/2) ;
            
			\coordinate (u000) at (0,0); 
			\coordinate (u100) at (\l/2,0); 
			\coordinate (u200) at (2*\l/2,0); 
                \coordinate (u010) at (\lx/2,\ly/2); 
                \coordinate (u110) at (\lx/2+\l/2,\ly/2); 
                \coordinate (u210) at (\lx/2+\l,\ly/2); 
                \coordinate (u020) at (\lx/2+\lx/2,\ly/2+\ly/2); 
                \coordinate (u120) at (\lx/2+\lx/2+\l/2,\ly/2+\ly/2); 
                \coordinate (u220) at (\lx/2+\lx/2+\l,\ly/2+\ly/2); 
	        \coordinate (u001) at (0,\l/2); 
			\coordinate (u101) at (\l/2,\l/2); 
			\coordinate (u201) at (2*\l/2,\l/2); 
                \coordinate (u011) at (\lx/2,\ly/2+\l/2); 
                \coordinate (u111) at (\lx/2+\l/2,\ly/2+\l/2); 
                \coordinate (u211) at (\lx/2+\l,\ly/2+\l/2); 
                \coordinate (u021) at (\lx/2+\lx/2,\ly/2+\ly/2+\l/2); 
                \coordinate (u121) at (\lx/2+\lx/2+\l/2,\ly/2+\ly/2+\l/2); 
                \coordinate (u221) at (\lx/2+\lx/2+\l,\ly/2+\ly/2+\l/2); 
                \coordinate (u002) at (0,\l); 
			\coordinate (u102) at (\l/2,\l); 
			\coordinate (u202) at (2*\l/2,\l); 
                \coordinate (u012) at (\lx/2,\ly/2+\l); 
                \coordinate (u112) at (\lx/2+\l/2,\ly/2+\l); 
                \coordinate (u212) at (\lx/2+\l,\ly/2+\l); 
                \coordinate (u022) at (\lx/2+\lx/2,\ly/2+\ly/2+\l); 
                \coordinate (u122) at (\lx/2+\lx/2+\l/2,\ly/2+\ly/2+\l); 
                \coordinate (u222) at (\lx/2+\lx/2+\l,\ly/2+\ly/2+\l); 

\draw[-] (u000)  --   (u100)--   (u200) ;
\draw[-] (u001)  --   (u101)--   (u201) ;
\draw[-] (u002)  --   (u102)--   (u202) ;


\draw[-] (u000)  --   (u001)--   (u002) ;
\draw[-] (u100)  --   (u101)--   (u102) ;
\draw[-] (u200)  --   (u201)--   (u202) ;

\draw[-] (u002)  --   (u012)--   (u022) ;
\draw[-] (u102)  --   (u112)--   (u122) ;
\draw[-] (u202)  --   (u212)--   (u222) ;
\draw[-] (u012)  --   (u112)--   (u212) ;
\draw[-] (u022)  --   (u122)--   (u222) ;

\draw[-] (u200)  --   (u210)--   (u220) ;
\draw[-] (u201)  --   (u211)--   (u221) ;
\draw[-] (u210)  --   (u211)--   (u212) ;
\draw[-] (u220)  --   (u221)--   (u222) ;
\draw[dashed] (u110)  --   (u111)--   (u112) ;
\draw[dashed] (u010)  --   (u011)--   (u012) ;
\draw[dashed] (u120)  --   (u121)--   (u122) ;
\draw[dashed] (u020)  --   (u021)--   (u022) ;
\draw[dashed] (u010)  --   (u110)--   (u210);
\draw[dashed] (u011)  --   (u111)--   (u211) ;
\draw[dashed] (u021)  --   (u121)--   (u221) ;
\draw[dashed] (u000)  --   (u010)--   (u020) ;
\draw[dashed] (u001)  --   (u011)--   (u021) ;
\draw[dashed] (u101)  --   (u111)--   (u121) ;
\draw[dashed] (u100)  --   (u110)--   (u120) ;
\draw[dashed] (u020)  --   (u120)--   (u220);

 \node[nodrt] (u220) at (\lx/2+\lx/2+\l,\ly/2+\ly/2) {}; 
\node[nod1] (u000) at (0,0) {};
\node[nod1] (u100) at (\l/2,0){};
\node[nod1] (u200) at (2*\l/2,0){};
\node[nod1] (u001) at (0,\l/2)  {};
\node[nod1] (u101) at (\l/2,\l/2)  {};
\node[nodt] (u201) at (2*\l/2,\l/2)  {};
\node[nod1] (u002) at (0,\l)  {};
\node[nodsw] (u011) at (\lx/2,\ly/2+\l/2)  {};
\node[nod1] (u010) at (\lx/2,\ly/2)  {};
\node[nodt] (u012) at (\lx/2,\ly/2+\l)  {};
\node[nod1] (u020) at (\lx/2+\lx/2,\ly/2+\ly/2) {};
\node[nod1] (u120) at (\lx/2+\lx/2+\l/2,\ly/2+\ly/2)  {};
\node[nod1] (u110) at (\lx/2+\l/2,\ly/2)  {};
\node[nod1] (u102) at (\l/2,\l)  {};
\node[nodrt] (u022) at (\lx/2+\lx/2,\ly/2+\ly/2+\l)   {};
\node[nodsw] (u111) at (\lx/2+\l/2,\ly/2+\l/2)  {};
\node[nodrt] (u202) at (2*\l/2,\l)  {};
\node[nod1] (u210) at (\lx/2+\l,\ly/2)  {};
\node[nodt] (u021) at (\lx/2+\lx/2,\ly/2+\ly/2+\l/2)   {};
\node[nodt] (u121) at (\lx/2+\lx/2+\l/2,\ly/2+\ly/2+\l/2)  {};
\node[nodt] (u211) at (\lx/2+\l,\ly/2+\l/2)  {};
\node[nodt] (u112)  at (\lx/2+\l/2,\ly/2+\l)  {};
\node[nodrt] (u212) at (\lx/2+\l,\ly/2+\l)  {};
\node[nodrt] (u122)  at (\lx/2+\lx/2+\l/2,\ly/2+\ly/2+\l)  {};
\node[nodrt] (u221) at (\lx/2+\lx/2+\l,\ly/2+\ly/2+\l/2)   {};
\node[noddb] (u222) at (\lx/2+\lx/2+\l,\ly/2+\ly/2+\l)   {};
		\end{tikzpicture}
		\caption{Consistency around a $3\times 3 \times 3$ cube for the lpBSQ equations \eqref{eq:qc}: {\bf step $0$}, the $12$ {\em black dots} represent the initial values; {\bf step $1$}, using the cubic  system, one gets unique values on the $2$ {\em white squares}; {\bf step} $2$, using the cubic   systems, one gets unique values on the $6$ {\em white triangles}; {\bf step} $3$, one has two ways to get values on the $6$ {\em reverse white triangles}, one is using the cubic system, the other  using the planar equation, this counts {\bf six} consistency checks; {\bf step} $4$, there are five ways to get the values on the {\em white diamond} that is the final point of the evolution, two are using the cubic systems, and three using the planar equations, this counts extra {\bf four} consistency checks. The consistency means all the values are compatible. } \label{fig:supercube-consistency}
	\end{figure}
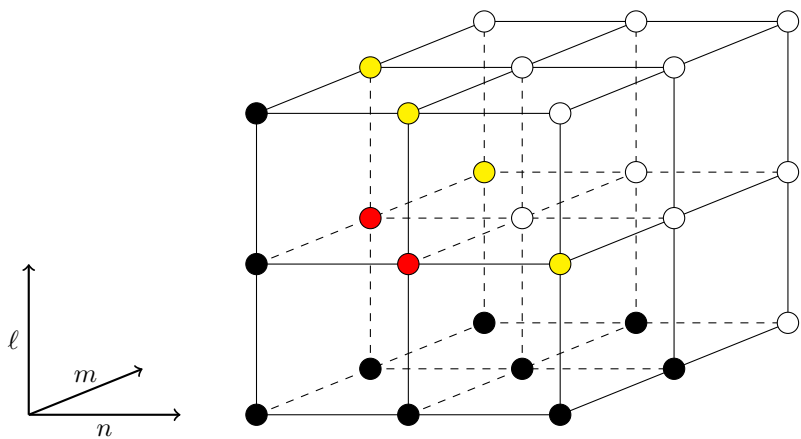
\begin{figure}[htb]  
\centering
		\begin{tikzpicture}[scale=1, decoration={markings,mark=at position 0.55 with {\arrow{latex}}}]
		\tikzstyle{nod}= [circle, inner sep=0pt, fill=white, minimum size=8pt, draw]
		\tikzstyle{nod1}= [circle, inner sep=0pt, fill=black, minimum size=8pt, draw]
            \tikzstyle{nod2}= [circle, inner sep=0pt, fill=red, minimum size=8pt, draw]
            \tikzstyle{nod3}= [circle, inner sep=0pt, fill=yellow, minimum size=8pt, draw]

			\def\lx{3}%
			\def\ly{1.22}%
			\def\lz{ (sqrt(\x*\x+\y*\y))}%
			\def\l{4}%
			\def\d{2}%
  \draw[thick,->] (-\d/2-\l/2,0) --node[below ]{$n$}  (-\d/2,0)  ;
    \draw[thick,->] (-\d/2-\l/2,0) --node[above ]{$m$}  (-\d/2-\l/2+\lx/2,0+\ly/2)  ;
        \draw[thick,->] (-\d/2-\l/2,0) --node[left ]{$\ell$}  (-\d/2-\l/2,\l/2) ;
            
			\coordinate (u000) at (0,0); 
			\coordinate (u100) at (\l/2,0); 
			\coordinate (u200) at (2*\l/2,0); 
                \coordinate (u010) at (\lx/2,\ly/2); 
                \coordinate (u110) at (\lx/2+\l/2,\ly/2); 
                \coordinate (u210) at (\lx/2+\l,\ly/2); 
                \coordinate (u020) at (\lx/2+\lx/2,\ly/2+\ly/2); 
                \coordinate (u120) at (\lx/2+\lx/2+\l/2,\ly/2+\ly/2); 
                \coordinate (u220) at (\lx/2+\lx/2+\l,\ly/2+\ly/2); 
	        \coordinate (u001) at (0,\l/2); 
			\coordinate (u101) at (\l/2,\l/2); 
			\coordinate (u201) at (2*\l/2,\l/2); 
                \coordinate (u011) at (\lx/2,\ly/2+\l/2); 
                \coordinate (u111) at (\lx/2+\l/2,\ly/2+\l/2); 
                \coordinate (u211) at (\lx/2+\l,\ly/2+\l/2); 
                \coordinate (u021) at (\lx/2+\lx/2,\ly/2+\ly/2+\l/2); 
                \coordinate (u121) at (\lx/2+\lx/2+\l/2,\ly/2+\ly/2+\l/2); 
                \coordinate (u221) at (\lx/2+\lx/2+\l,\ly/2+\ly/2+\l/2); 
                \coordinate (u002) at (0,\l); 
			\coordinate (u102) at (\l/2,\l); 
			\coordinate (u202) at (2*\l/2,\l); 
                \coordinate (u012) at (\lx/2,\ly/2+\l); 
                \coordinate (u112) at (\lx/2+\l/2,\ly/2+\l); 
                \coordinate (u212) at (\lx/2+\l,\ly/2+\l); 
                \coordinate (u022) at (\lx/2+\lx/2,\ly/2+\ly/2+\l); 
                \coordinate (u122) at (\lx/2+\lx/2+\l/2,\ly/2+\ly/2+\l); 
                \coordinate (u222) at (\lx/2+\lx/2+\l,\ly/2+\ly/2+\l); 

\draw[-] (u000)  --   (u100)--   (u200) ;
\draw[-] (u001)  --   (u101)--   (u201) ;
\draw[-] (u002)  --   (u102)--   (u202) ;


\draw[-] (u000)  --   (u001)--   (u002) ;
\draw[-] (u100)  --   (u101)--   (u102) ;
\draw[-] (u200)  --   (u201)--   (u202) ;

\draw[-] (u002)  --   (u012)--   (u022) ;
\draw[-] (u102)  --   (u112)--   (u122) ;
\draw[-] (u202)  --   (u212)--   (u222) ;
\draw[-] (u012)  --   (u112)--   (u212) ;
\draw[-] (u022)  --   (u122)--   (u222) ;

\draw[-] (u200)  --   (u210)--   (u220) ;
\draw[-] (u201)  --   (u211)--   (u221) ;
\draw[-] (u210)  --   (u211)--   (u212) ;
\draw[-] (u220)  --   (u221)--   (u222) ;
\draw[dashed] (u110)  --   (u111)--   (u112) ;
\draw[dashed] (u010)  --   (u011)--   (u012) ;
\draw[dashed] (u120)  --   (u121)--   (u122) ;
\draw[dashed] (u020)  --   (u021)--   (u022) ;
\draw[dashed] (u010)  --   (u110)--   (u210);
\draw[dashed] (u011)  --   (u111)--   (u211) ;
\draw[dashed] (u021)  --   (u121)--   (u221) ;
\draw[dashed] (u000)  --   (u010)--   (u020) ;
\draw[dashed] (u001)  --   (u011)--   (u021) ;
\draw[dashed] (u101)  --   (u111)--   (u121) ;
\draw[dashed] (u100)  --   (u110)--   (u120) ;
\draw[dashed] (u020)  --   (u120)--   (u220);

                \node[nod1] (u000) at (0,0) 
                {};
                \node[nod1] (u100) at (\l/2,0) 
                {};
                \node[nod1] (u200) at (2*\l/2,0) 
                {};             
                \node[nod1] (u010) at (\lx/2,\ly/2)  {};
                \node[nod1] (u110) at (\lx/2+\l/2,\ly/2)  {};
                \node[nod1] (u210) at (\lx/2+\l,\ly/2)  {};
                \node[nod1] (u020) at (\lx/2+\lx/2,\ly/2+\ly/2) 
                {};
                \node[nod1] (u120) at (\lx/2+\lx/2+\l/2,\ly/2+\ly/2)  {};
                \node[nod] (u220) at (\lx/2+\lx/2+\l,\ly/2+\ly/2)
                {};
                \node[nod1] (u001) at (0,\l/2)  {};
                \node[nod2] (u101) at (\l/2,\l/2)  {};
                \node[nod3] (u201) at (2*\l/2,\l/2)  {};
                \node[nod2] (u011) at (\lx/2,\ly/2+\l/2)  {};
                \node[nod] (u111) at (\lx/2+\l/2,\ly/2+\l/2)  {};
                \node[nod] (u211) at (\lx/2+\l,\ly/2+\l/2)  {};
                \node[nod3] (u021) at (\lx/2+\lx/2,\ly/2+\ly/2+\l/2)   {};
                \node[nod] (u121) at (\lx/2+\lx/2+\l/2,\ly/2+\ly/2+\l/2)  {};
                \node[nod] (u221) at (\lx/2+\lx/2+\l,\ly/2+\ly/2+\l/2)   {};
                \node[nod1] (u002) at (0,\l)  {};
                \node[nod3] (u102) at (\l/2,\l)  {};
                \node[nod] (u202) at (2*\l/2,\l)  {};
                \node[nod3] (u012) at (\lx/2,\ly/2+\l)  {};
                \node[nod] (u112)  at (\lx/2+\l/2,\ly/2+\l)  {};
                \node[nod] (u212) at (\lx/2+\l,\ly/2+\l)  {};
                \node[nod] (u022) at (\lx/2+\lx/2,\ly/2+\ly/2+\l)   {};
                \node[nod] (u122)  at (\lx/2+\lx/2+\l/2,\ly/2+\ly/2+\l)  {};
                \node[nod] (u222) at (\lx/2+\lx/2+\l,\ly/2+\ly/2+\l)   {};
		\end{tikzpicture}
		\caption{The initial-value problem for the lpBSQ equations \eqref{eq:qc} is well-defined with the following choices of initial data: in addition to assigning values to the $10$ {\em black dots}, assigning one value to any one of the two {\em red dots} and one valuue to any one of the four {\em yellow dots}. Then,  the lpBSQ equations \eqref{eq:qc} is consistent around the $3\times 3\times 3$ cube.} \label{fig:supercube-initial-values}
	\end{figure}

The evolutionary-type equations allow us to set up an initial-value problem on a $2\times 2 \times 2$ cube. This is explained in Figure~\ref{fig:supercube-consistency}:  besides the lattice parameters, initial data are assigned to $12$ vertices among $27$ vertices on a $2\times 2\times 2$ cube with values to be determined on $15$ vertices. There are in total $9+2\times 8 =25$ equations: $9$ nine-point equations, $2$ 3D companions of lpKdV on an elementary cube as explained in Remark~\ref{rm:36} with $8$ elementary cubes in a $2\times 2\times 2$ cube. Therefore, there are in total $25-15=10$  consistency checks. This allows us to state the following results which can be seen as a generalization of the 3D consistency property for the lpBSQ equations \eqref{eq:qc}.
\begin{theorem}
The lpBSQ equations \eqref{eq:qc} are consistent around  a $2\times 2 \times 2$ cube.  
\end{theorem}
The consistency can be directly check.     
Finally, let us comment on the combinatorial aspect in assigning the initial values on a $3\times 3\times 3$ cube. There is a freedom to assign initial values on the cube as explained in Figure~\ref{fig:supercube-initial-values}.

\section{Lattice BSQ-Q3 system}\label{sec:4}
In this section, we provide the derivation of the lattice BSQ-Q3 system \eqref{eq:fuvz}.
The parameter $\delta$ arises from a $GL_3$ action, in direct analogy with the $GL_2$ origin of 
$\delta$ in the KdV–Q3 case. We prove the 3D consistency of the system and show that its degeneration 
$r=0$ yields a $PGL_3$-invariant lattice system, which generalises the classical $PGL_2$-invariant Schwarzian BSQ equation. 
\subsection{Derivations of  lattice BSQ-Q3 and 3D consistency}
The setting up of the derivation is similar to that of lattice KdV-Q3 as presented in Section \ref{sec:22}. It relies on a gauge transformation between two Lax pairs of lpBSQ with different
spectral parameters.  We follow the notations introduced in Section \ref{sec:3}. 
The derivation  is split into the following steps.
For simplicity, set $\ell = 0$. 
\begin{description}
    \item[Step 1: discrete gauge transformation.]

 Let $U_q$ and $ V_q$ be another pair of Lax matrices by switching the spectral parameter $p$ to $q$, \ie  $U_q = U_p\vert_{p\to q}$ and $     V_q = V_p\vert_{p\to q}$, and let $\Phi_q$ be the fundamental solution of 
\begin{equation}
  \label{eq:BSQUV1}
    U_q\Phi_q=\wt{\Phi}_q\,,\quad     V_q\Phi_q =\wh{\Phi}_q\,, 
\end{equation}
normalized similarly as \eqref{eq:bsqn}. Clearly, compatibility condition of \eqref{eq:BSQUV1} again leads to the lpBSQ system \eqref{eq:3csyst}. Let $G$ be a $3\times 3$ gauge matrix connecting $\Phi_p$ and $\Phi_q$ such as
\begin{equation}
\label{eq:gpq}
    G \,\Phi_p = \Phi_q\,,
\end{equation}
then the pairs $U_p, V_p$ and $U_q, V_q$ are connected by 
    \begin{equation}\label{eq:gbsq}
        \wt{G}\, U_p = U_q\,G\,,\quad         \wh{G}\, V_p = V_q\,G\,. 
    \end{equation}
Due to the normalization \eqref{eq:bsqn}, 
\begin{equation}\label{eq:gbsq1}
    \det G\vert_{\ell  = 0} =  \frac{(q^3 -\alpha^3  )^n(q^3 -\beta^3)^m }{(p^3 -\alpha^3)^n(p^3 - \beta^3)^m}\,. 
\end{equation}

    \item[Step 2: assignment of the variables.] The BSQ-Q3 variable $u, v, z$ are defined as follows 
    \begin{equation}\label{eq:bsquvz}
    u =  \text{tr}(G)\,,\quad    v  =  \frac{1}{2}(( \text{tr}(G))^2 -\tr(G^2))  \,,\quad z  = \frac{G^{31}}{g_{13}} 
    \end{equation}
  where  $ g_{ij}= (G)_{ij}$ which is the $ij$-entry of $G$ and $G^{ij}$ is the $ij$-minor of $G$. In terms of components of $G$, one has explicitly 
  \begin{subequations}\label{eq:bsquv}
\begin{align}\label{eq:bsquv1}
      u =  & g_{11}+g_{22}+g_{33}\,, \\    \label{eq:bsquv2}v  = & g_{11} g_{22} + g_{11} g_{33} + g_{22} g_{33} -g_{12} g_{21}  - g_{13} g_{31} - g_{23} g_{32} \,,    \\
\label{eq:bsquv3}      z = & \frac{1}{g_{13}}\begin{vmatrix}
        g_{12} & g_{13} \\ g_{22} &  g_{23}
    \end{vmatrix} = \frac{g_{12} g_{23} -g_{13}g_{22}}{g_{13}}\,.
\end{align}  \end{subequations}
Note that the variable $v$ can also be expressed as 
\begin{equation}
    v = \det G\,\tr( G^{-1})\,.
\end{equation}
The introduction of these variables forms the key to the construction of the BSQ-Q3 system generalising the KdV case \eqref{eq:utrg}. Here, $u, v$ are the direct generalisations of the KdV-Q3 variable, \cf \eqref{eq:utrg}, to the $3\times 3 $ gauge transformation case \eqref{eq:gbsq}. The variable $z$, will 
   \item[Step 3: shift of the variables.] Take (\ref{eq:gbsq}-\ref{eq:bsquvz}) as the governing systems: on the one hand, solving the
linear systems \eqref{eq:gbsq} one can get ~$\wt{~}$~ and ~$\wh{~}$~ shifts of $g_{ij}$ as expressions of the remaining
variables that are $g_{ij}$,  lpBSQ variables $w,x,y$ and their shifts; on the other, using \eqref{eq:bsquv1}, \eqref{eq:bsquv2} and \eqref{eq:gbsq1}, one can get $g_{11}, g_{21}, g_{31}$ as expressions of $u,v$ and remaining $g_{ij}$. The origin of the form of  $z $ is less transparent than that of $u$ and  $v$. This specific combination is uniquely selected by the condition that the evolution equations \eqref{eq:fuvz} form a closed system on the square lattice.

   \begin{description}
       \item[Step 3.1: shifts of $z$.] It is straightforward to make  ~$\wt{~}$~ and ~$\wh{~}$~ shifts of $z$ using the substitutions explained above. This leads to \eqref{eq:tz} and \eqref{eq:hz} respectively. 
       \item[Step 3.2: shifts of $u,v$.]   Recall $f_u, g_u, f_v, g_v$ defined in \eqref{eq:fguv}, this allows us to compute ~$\wt{~}$~ and ~$\wh{~}$~ shifts of $u$ and $v$. One gets
\begin{equation}\label{eq:bsqfg}
\frac{1}{p^3 -q^3}\bma f_u \\ f_v \ema   =   
\cG \bma \wt{w} \\ \wt{x}\ema -\cF\,,\quad \frac{1}{p^3 -q^3}\bma g_u \\ g_v \ema   =   
\cG \bma \wh{w} \\ \wh{x}\ema -\cF\,,
\end{equation}
  where
  \begin{equation}
      \cG = \bma  g_{12} & g_{13} \\ G^{21} & -G^{31} \ema \,, \quad \cF = \bma g_{22} +g_{33} \\ G^{11} \ema 
  \end{equation}

   \end{description} 
 
    \item [Step 4: Miura-type transformations.] Now let us describe the Miura-type transformations.  
    \begin{description}
        \item[Step 4.1: $\wh{w}-\wt{w}$ and $\wh{x}-\wt{x}$.]       It follows from \eqref{eq:bsqfg} that $\wt{w}, \wh{w}, \wt{x}, \wh{x}$ can be expressed in terms of $f_u,f_v, g_u,g_v$ and $g_{ij}$'s.  In particular, one can get
\begin{equation}\label{eq:dwdx}
    \wh{w}-\wt{w} =  \frac{G^{31}(g_u-f_u) +g_{13}(g_v-f_v)}{(p^3-q^3)(g_{12} G^{31} + g_{13} G^{21})}\,,\quad
              \wh{x}-\wt{x} =  \frac{              G^{21}(g_u-f_u) -g_{12}(g_v -f_v)}{(p^3-q^3)(g_{12} G^{31} + g_{13} G^{21})}\,.
\end{equation}
    \item[Step 4.2:  $\wh{\wt{w}}, \wh{\wt{x}}$ and $\wt{\wh{w}}, \wt{\wh{x}}$.] Taking  ~$\wh{~}$~ shift of the left equation of  \eqref{eq:bsqfg}, this allows us to compute $\wh{\wt{w}}$ and $\wh{\wt{x}}$ in terms of $\wh{\wt{u}}$ and $\wh{\wt{v}}$  (contained in $\wh{f}_u$ and $ \wh{f}_v$ as given in \eqref{eq:fguv}) and remaining variables (using again the substitutions explained in Step 4.1 for $\wt{w}, \wh{w}, \wt{x}, \wh{x}$). We omit the explicit expressions here. Inserting \eqref{eq:dwdx} and expressions of $\wh{\wt{w}}, \wh{\wt{x}}$ into \eqref{eq:3csyst1} and \eqref{eq:3csyst3}, one can get a system for $\wh{\wt{u}}$ and $\wh{\wt{v}}$, which  leads to  \eqref{eq:thuu} and \eqref{eq:thv} with $\delta=1$.
         Similarly, take  ~$\wt{~}$~ shift of the right equation of \eqref{eq:bsqfg}, one uses $\wt{\wh{u}}, \wt{\wh{v}}$ to denote ~$\wt{~}$~ of $\wh{u}, \wh{v}$. Then, one can get again 
         \eqref{eq:thuu} and \eqref{eq:thv} with $\delta=1$ by setting $\wt{\wh{u}} =\wh{\wt{u}}$ and $\wt{\wh{v}} =\wh{\wt{v}}$ following the procedure explained above. Lastly, we can also check that the above Miura-type transformations is consistent with \eqref{eq:3csyst1} and \eqref{eq:3csyst3} in the lpBSQ system.  
    \end{description}

    \item[Step 5:  $\delta$ term induced by $GL_3$ action.]
      Under an action of a $3\times 3$ constant matrix $M$ (not necessarily nondegenerate) on \eqref{eq:BSQUV1} as
  \begin{equation}\label{eq:g3a}
      \Phi_q \mapsto \Phi_q^M =  \Phi_q\,M\,, 
  \end{equation}   
$\Phi_q^M$ is again a solution of \eqref{eq:KDVUV1},  and using 
$  G\, \Phi_p  = \Phi_q^M$, the above steps  still hold. This action induces 
\begin{equation}
    \det G  = \det M \, \frac{(q^3 - \alpha^3  )^n(q^3 - \beta^3)^m }{(p^3 -\alpha^3  )^n(p^3 -\beta^3)^m}\,.  
\end{equation}Let $\det M = \delta$, one gets precisely the lattice BSQ-Q3 system \eqref{eq:fuvz}. Therefore, when $\delta \neq 0$, the $\delta$ term  is a result of the $GL_3$ action. 
\end{description}

   In view of the assignments \eqref{eq:bsquvz}, and using $G = \Phi_q^M \Phi_p^{-1}$ with $\Phi_q^M$ given in \eqref{eq:g3a}, one has respectively the expressions of $u,v,z$ in terms of components of $\Phi_p$ and $\Phi_q$. Let  $ \phi_{p,ij}$ and  $ \phi_{q,ij}$ be the $ij$-entry of $\Phi_p$ and $\Phi_q$ respectively, and  let $\Phi^{ij}_p$ and $\Phi^{ij}_q$ be the $ij$-minor of $\Phi_p$ and  $\Phi_q$ respectively. Also let $ M^{ij}$ be the $ij$-minor of $M$, $m_{ij}$ be the $ij$-entry of $M$. Denote $\Phi_p$ and $\Phi_q$ by   \begin{equation}
    \Phi_p = \bma \Phi^1_p & \Phi^2_p & \Phi^3_p \ema \,,\quad  \Phi_q = \bma \Phi^1_q & \Phi^2_q & \Phi^3_q \ema \,.
\end{equation} Then, 
\begin{subequations}
\label{eq:eig}\begin{align}
    u =& \frac{\sum_{j=1}^3\left( m_{j1} \begin{vmatrix}
        \Phi^j_q &\Phi^2_p  &\Phi^3_p 
    \end{vmatrix}+m_{j2} \begin{vmatrix}
        \Phi^j_q &\Phi^3_p  &\Phi^1_p 
    \end{vmatrix}+m_{j3} \begin{vmatrix}
        \Phi^j_q &\Phi^1_p  &\Phi^2_p 
    \end{vmatrix}\right)}{(p^3 -\alpha^3  )^n(p^3 -\beta^3)^m }\,, \\
    v =& \frac{\sum_{j=1}^3\, (-1)^{1+j}\left(M^{1j} \begin{vmatrix}
        \Phi^j_p &\Phi^2_q  &\Phi^3_q 
    \end{vmatrix}+M^{2j} \begin{vmatrix}
        \Phi^j_p &\Phi^3_q  &\Phi^1_q 
    \end{vmatrix}+M^{3j} \begin{vmatrix}
        \Phi^j_p &\Phi^1_q  &\Phi^2_q 
    \end{vmatrix}\right)}{(p^3 -\alpha^3  )^n(p^3 -\beta^3)^m }\,,\\
    z = &\frac{\sum_{j=1}^3\left( - M ^{j1}\,\Phi_q^{3j}\, \phi_{p,11}+M ^{j2}\,\Phi_q^{3j}\,  \phi_{p,12}-M ^{j3}\,\Phi_q^{3j}\,  \phi_{p,13}\right)}{\sum_{j=1}^3\left( - m_{j1}\,\Phi_{p}^{31}\,\phi_{q,1j}+m_{j2}\,  \Phi_{p}^{32}\,\phi_{q,1j}-m_{j3}\,  \Phi_{p}^{33}\,\phi_{q,1j}\right)}\,.
 \end{align}
\end{subequations}
In other words, $u,v,z$ can be expressed as combinations of the eigenfunctions. 

The lattice BSQ-Q3 system admits an autonomous version. Consider the set of parameters $s,\sigma, t,\tau$ introduced in \eqref{sstt}. Fix the branches of the cubic roots such that standard exponent rules apply, in particular, 
\begin{equation}
  \Lambda: = \frac{\sigma^n \tau^m}{s^n t^m} \,,\quad    \Lambda^2 = \frac{\sigma^{2n} \tau^{2m}}{s^{2n} t^{2m}}\,,\quad \Delta = \Lambda^3 \delta\,. 
\end{equation}
Introduce a new set of dynamical variables $\chi, \xi, \zeta$ related to $u,v,z$ respectively through
\begin{equation}
  \chi = u \Lambda^{-1}\,,\quad \xi = v \Lambda^{-2}\,,\quad \zeta = z \Lambda^{-1}\,. 
\end{equation}
It follows from the above transformation that a set of functions  \begin{equation}
\ff_u  =  \sigma^3 \chi -s^2 \sigma \wt{\chi}\,,\quad \ff_v =\sigma^3 \xi -s \sigma^2 \wt{\xi} \,,\quad \fg_u  =  \tau^3 \chi -t^2 \tau \wh{\chi}\,,\quad \fg_v =\tau^3 \xi -t \tau^2 \wh{\xi}\,, 
\end{equation}
can be defined related to those introduced in  \eqref{eq:fguv} through
\begin{equation}
f_u = \Lambda \ff_u\,,\quad   g_u = \Lambda \fg_u\,,\quad      f_v = \Lambda^2 \ff_v\,,\quad     g_v = \Lambda^2 \fg_v\,. 
\end{equation}
Then the non-autonomous lattice BSQ-Q3 system \eqref{eq:fuvz} can be written as (also letting $\delta \to \frac{st \delta}{r}$)
\begin{subequations}\label{eq:autouvz}
\begin{align}
  \wh{\wt{\chi}} =&\frac{(st(\sigma t \wt{\chi}\fg_u -s\tau \wh{\chi}\ff_u)+r(s^2\tau^2 \wt{\xi} -\sigma^2t^2 \wh{\xi}) )\zeta
                    + s t (\sigma t \wt{\chi}\fg_v - s\tau \wh{\chi}\ff_v)
                    +\rho r \delta}{
                    st\sigma \tau ((\fg_u-\ff_u)\zeta +\fg_v-\ff_v)    },\\
  \wh{\wt{\xi}} =&\frac{ ( \sigma^2\tau^2 (s \sigma \wh{\xi} (\sigma\wt{\chi} -s\chi) - t\tau \wt{\xi}(\tau \wh{\chi}-t\chi))+\rho r \delta          )\zeta +\sigma^2t^2 \wt{\xi}\fg_v-s^2\tau^2 \wh{\xi}\ff_v+(\tau^3 \ff_u-\sigma^3\fg_u)\delta
}{                \sigma^2 \tau^2 ((\fg_u-\ff_u)\zeta +\fg_v-\ff_v)    }, 
\end{align}
together with the side equations
\begin{align}
  \frac{\sigma}{s}\wt{\zeta}=& \frac{( \ff_u\ff_v+r\chi \ff_v  + st\, r \delta)\zeta + \ff_v^2 - st \ff_u\delta+ r\xi \ff_v    }{(-\ff_u^2-r \ff_v-r\chi\ff_u -r^2\xi)\zeta     -  ( \ff_u\ff_v+r\chi \ff_v  + st\,r \delta)} \,, \\
    \frac{\tau}{t}\wh{\zeta}=& \frac{( \ff_u\ff_v+r\chi \ff_v  + st\, r \delta)\zeta + \ff_v^2 - st \ff_u\delta+ r\xi \ff_v    }{(-\ff_u^2-r \ff_v-r\chi\ff_u -r^2\xi)\zeta     -  ( \ff_u\ff_v+r\chi \ff_v  + st\,r \delta)} \,.
\end{align}
\end{subequations}

 The lattice BSQ-Q3 system \eqref{eq:fuvz} is by construction consistent around an elementary cube. To check the consistency, one needs to transform it into an evolutionary form. This can be done by taking the ~$\wh{~}$~ shift of \eqref{eq:tz} with substitutions of $\wh{\wt{u}}$ and $\wh{\wt{v}}$ using \eqref{eq:thuu} and \eqref{eq:thv}. Then, one can get the expression of $\wt{\wh{z}}$ 
\begin{equation}\label{eq:thzz}
    \wt{\wh{z}}=\frac{ (\tau^3f_v -\sigma^3g_v)(f_v g_u - g_v f_u +r((f_v - g_v)u- (f_u - g_u)v))
      + E_1}{
      s^3g_u(f_vg_u  - f_u g_v) +t^3 f_u (f_u g_v - f_v g_u)
    +  r^2 \rho (g_u-f_u)\delta \Delta+E_2 }\,. 
\end{equation}
where
\begin{equation}
E_1=  (r (\sigma g_u-\tau f_u)(f_u-g_u) + r^2 \rho (g_v - f_v)\delta \Delta\,,~~ E_2= r\rho u (f_u g_v -f_v g_u  )  +r s^3t^3 (\wh{v} - \wt{v}) (f_v - g_v) \,. 
\end{equation}
Alternatively, taking the ~$\wt{~}$~ shift of \eqref{eq:hz} with substitutions of $\wh{\wt{u}}$ and $\wh{\wt{v}}$ using \eqref{eq:thuu} and \eqref{eq:thv} leads to the same result. This also ensures an internal consistency of \eqref{eq:fuvz} \ie $\wt{\wh{z}} = \wh{\wt{z}}$.

Note that the expression of $\wh{\wt{z}}$ is independent of $z, \wt{z}, \wh{z}$. This is similar to $\wt{\wh{y}}$ in the evolutionary form of the three-component lpBSQ system as showed in \eqref{eq:thbwxy}. 
 Now, on an elementary square lattice, having $(u,\wt{u}, \wh{u}, v,\wt{v},\wh{v}, x)$ as well as the lattice parameters as initial data, one can get the values of $\wh{\wt{u}}, \wh{\wt{v}}, \wh{\wt{z}}$ using \eqref{eq:thuu}, \eqref{eq:thv} and \eqref{eq:thzz} respectively. The values of $\wt{z}$ and $\wh{z}$ can be obtained using \eqref{eq:tz} and \eqref{eq:hz}. This allows us to set up an initial-value problem on an elementary cube. Then, one can check directly the 3D consistency property. 

\begin{theorem}
The lattice BSQ-Q3 system \eqref{eq:fuvz} is  consistent around a cube. 
\end{theorem}

\subsection{$PGL_3$-invariant BSQ system}
An obvious degeneration of the system is to set $\delta = 0$ so that the terms involving $\Delta$ vanish in \eqref{eq:fuvz}. The resulting  system is referred to as lattice BSQ-Q3$\vert_{\delta=0}$.

Another straight degeneration is to set $r = 0$, or $p^3 = q^3$. In this case, \eqref{eq:fuvz}  reduces to 
\begin{subequations}\label{eq:sl3zuv}
\begin{equation}\label{eq:zuv}
    \wt{z} ( \wt{u}-  u) = v-\wt{v}\,,\quad \wh{z} (\wh{u}  - u) = v-\wh{v}\,,
\end{equation}
and
    \begin{align}
\label{eq:sl3u}        \wh{\wt{u}} & =\frac{(\alpha^3 - p^3) \wh{u} (\wt{v}- v + \wt{u} z - u z) - (\beta^3 - p^3) \wt{u} (\wh{v} - v + \wh{u} z - u z)}{(\alpha^3 - p^3) (\wt{v} - v + \wt{u} z - u z) - (\beta^3 - p^3) (z)}\,, \\
\label{eq:sl3v}                \wh{\wt{v}} & =\frac{(\alpha^3 - p^3) \wh{v} (\wt{v}- v + \wt{u} z - u z) - (\beta^3 - p^3) \wt{v} (\wh{v} - v + \wh{u} z - u z)}{(\alpha^3 - p^3) (z) - (\beta^3 - p^3) (\wh{v}- v + \wh{u} z - u z)} \,.
    \end{align}
\end{subequations}
This equation is 3D-consistent and the internal consistency can be easily checked: compute respectively $\wh{\wt{z}}$ and  $\wt{\wh{z}}$ as ~$\wh{~}$~ and   ~$\wt{~}$~ shifts of the first and second equations in \eqref{eq:zuv}, they coincides by taking account of both \eqref{eq:sl3u} and \eqref{eq:sl3v}. In \eqref{eq:sl3zuv}, $u $ and $v$ play symmetric roles. Moreover, one can eliminate the variable $z$ in the system and get two coupled systems for $u,v$:
\begin{subequations}\label{sl3-invar}
    \begin{align}
    \frac{\wt{u}-\wh{\wt{u}}}{\wh{u}-\wh{\wt{u}}} =     \frac{\wt{v}-\wh{\wt{v}}}{\wh{v}-\wh{\wt{v}}} =&\frac{\alpha^3-p^3}{\beta^3-p^3}
    \begin{vmatrix}
        1 & 1 & 1 \\ u  & \wt{u} & \underset{\widetilde{~~}}{u}\\v & \wt{v} & \underset{\widetilde{~~}}{v}
    \end{vmatrix}\begin{vmatrix}
        1 & 1 & 1 \\ u  & \wh{u} & \underset{\widetilde{~~}}{u}\\v & \wh{v} & \underset{\widetilde{~~}}{v}
    \end{vmatrix}^{-1}\,, \\
\frac{\wh{u}-\wh{\wt{u}}}{\wt{u}-\wh{\wt{u}}} =     \frac{\wh{v}-\wh{\wt{v}}}{\wt{v}-\wh{\wt{v}}} =&\frac{\beta^3-p^3}{\alpha^3-p^3}
    \begin{vmatrix}
        1 & 1 & 1 \\ u  & \wh{u} & \underset{\widehat{~~}}{u}\\v & \wh{v} & \underset{\widehat{~~}}{v}
    \end{vmatrix}\begin{vmatrix}
        1 & 1 & 1 \\ u  & \wt{u} & \underset{\widehat{~~}}{u}\\v & \wt{v} & \underset{\widehat{~~}}{v}
    \end{vmatrix}^{-1}\,. 
\end{align}
\end{subequations}
Each of the above systems is defined on a five-point stencil, and one can be obtained the other by interchanging $\{~\wt{~}~,\alpha\} $ and $\{~\wh{~}~,\beta\} $.   
These systems possess a remarkable property that they are invariant under $PGL_3$ action:
\begin{equation}
     (u, v) \mapsto (\frac{m_{11} u +m_{12}v+m_{13}}{m_{31} u +m_{32}v+m_{33}}, \frac{m_{21} u +m_{22}v+m_{23}}{m_{31} u +m_{32}v+m_{33}} )
\end{equation}
where $m_{ij}  = (M)_{ij} $ is the $ij$-entry of a $3\times 3$ nondegenerate matrix. A detailed study of  $PGL_3$-invariant integrable system and related integrable system from the perspective of factorization of linear operators will appear in \cite{NPZZ}. One can further represent the system \eqref{eq:sl3zuv} in the form
\begin{subequations}
\begin{align}
        \frac{(\alpha^3-p^3)(\wt{z}-z)}{(\beta^3-p^3)(\wh{z}-z)} &= \frac{(\wh{\wt{u}}- \wt{u}) (\wh{u} - u)}{(\wh{\wt{u}} - \wh{u}) (\wt{u} - u)} \,,\\
                \frac{(\alpha^3-p^3)(\wt{z}-z)\wh{z}}{(\beta^3-p^3)(\wh{z}-z)\wt{z}} &= \frac{(\wh{\wt{v}}- \wt{v}) (\wh{v} - v)}{(\wh{\wt{v}} - \wh{v}) (\wt{v} -v)} \,.
\end{align}    
\end{subequations}
This amount to $2$ two-components system on an elementary square lattice as
\begin{subequations}
\begin{equation}\label{eq:uz}
       \frac{(\alpha^3-p^3)(\wt{z}-z)}{(\beta^3-p^3)(\wh{z}-z)} = \frac{(\wh{\wt{u}}- \wt{u}) (\wh{u} - u)}{(\wh{\wt{u}} - \wh{u}) (\wt{u} - u)}\,,\quad \wh{\wt{z}} = \frac{\wh{z}(\wh{u}-u) -\wt{z}(\wt{u}-u)  }{\wh{u}-\wt{u}}\,,
\end{equation}
and
\begin{equation}
      \frac{(\alpha^3-p^3)(\wt{z}-z)\wh{z}}{(\beta^3-p^3)(\wh{z}-z)\wt{z}} = \frac{(\wh{\wt{v}}- \wt{v}) (\wh{v} - v)}{(\wh{\wt{v}} - \wh{v}) (\wt{v} -v)} \,,\quad \wh{\wt{z}} = \frac{\wt{z}\wh{z} (\wh{v} - \wt{v})   }{\wt{z}(\wh{v}-v) -\wh{z}(\wt{v}-v) }\,.
\end{equation}
\end{subequations}
One can further eliminate the variables $z$  to get a single-component equation for $u$ or $v$ defined on a nine-point  square lattice: let 
\begin{equation}
  { \bf z}  = \wt{z}-z\,,\quad  {\bf s}  
 = \wh{z}-z\,,\quad     \Gamma(u): = \frac{(\wh{\wt{u}}- \wt{u}) (\wh{u} - u)}{(\wh{\wt{u}} - \wh{u}) (\wt{u} - u)}\,,
\end{equation}
where $\Gamma(u)$ is a cross-ratio of $u,\wt{u},\wh{u}, \wh{\wt{u}}$. 
It follows from  \eqref{eq:uz} that\begin{equation}
     \frac{\alpha^3 -p^3}{\beta^3 - p^3}\frac{\bf z}{ {\bf s}} = \Gamma(u)\,, \quad \wh{u}\,\wh{{\bf z}}
    -\wt{u} \, \wt{{\bf s}} =  u \,{\bf z} -u \,{\bf s}\,,
\end{equation}
then, together with the relation $
    { \bf z} - {\bf s} = \wh{{\bf z}} - \wt{{\bf s}}$, 
one gets 
\begin{equation}
    \wh{\bf z} = \frac{({\bf z} -{\bf s} )(\wt{u}-u)}{(\wt{u}-\wh{u})}\,,\quad 
    \wt{{\bf s}} = \frac{({\bf z} -{\bf s} )(\wh{u}-u)}{(\wt{u}-\wh{u})}\,. 
\end{equation}
On the one hand,  one has
\begin{equation}
   \frac{\beta^3 -p^3}{\alpha^3 - p^3} \,\wh{\wt{\Gamma}}(u) = \frac{\wh{\wt{\bf z}}}{\wh{\wt{\bf s}}} =\frac{(\wt{\bf z} - \wt{\bf s} ) (\wh{\wh{u}} -\wt{\wh{u}} ) (\wt{\wt{u}}  -\wt{u})}{(\wh{\bf z} - \wh{\bf s})(\wt{\wh{u}} - \wt{\wt{u}}) (\wh{\wh{u}} - \wh{u}) }\,, 
\end{equation}
on the other hand
\begin{equation}
    \frac{\wt{\bf z} - \wt{\bf s}}{\wh{\bf z} - \wh{\bf s} } = \frac{\wt{\bf s} }{\wh{\bf z} } \,  \frac{(\beta^3 -p^3)}{(\alpha^3 - p^3)} \,\frac{ (\alpha^3 - p^3) - (\beta^3 -p^3) \wt{\Gamma}(u)}{(\alpha^3 - p^3) - (\beta^3 -p^3) \wh{\Gamma}(u)}\,. 
\end{equation}
Combining the above two equations together amounts to the  lattice-Schwarzian BSQ equations \cite{Weiss} which is $PGL_2$-invariant. 
 \begin{equation}\label{eq:sl2bsq}
     \frac{(\wh{\wh{\wt{\wt{u}}}} - \wh{\wt{\wt{u}}} ) (\wh{\wt{u}} - \wt{\wt{u}}) (\wt{u} - u)}{ (\wh{\wh{\wt{\wt{u}}}} - \wh{\wh{\wt{u}}})(\wh{\wh{u}} - \wh{\wt{u}}) (\wh{u} - u)}=
     \frac{(\alpha^3-p^3)( \wh{\wt{\wt{u}}}-\wh{\wt{u}}   ) ( \wt{\wt{u}} - \wt{u})-(\beta^3-p^3)(\wh{\wt{\wt{u}}} - \wt{\wt{u}}) (\wh{\wt{u}}  - \wt{u})}{(\alpha^3-p^3)( \wh{\wh{\wt{u}}} -\wh{\wh{u}}) (\wh{\wt{u}} - \wh{u})-(\beta^3-p^3)(\wh{\wh{\wt{u}}} -\wh{\wt{u}} )(\wh{\wh{u}}  - \wh{u}) }\,.
 \end{equation}

In analogy with the ABS classification, the degenerate system \eqref{eq:sl3zuv} may be regarded as the BSQ counterpart of the Q1$_{\delta=0}$  equation\textemdash also known as the lattice Schwarzian KdV, or cross-ratio equation, which plays an important role in the theory of integrable lattice equations. Moreover, this system exhibits an implicit connection to the pentagram maps studied in \cite{OST}, as its continuous limit coincides with the continuous limit of those maps \footnote{This correspondence will be explained in  \cite{NPZZ}.}.
\section{Conclusion}

In this paper, we establish two new results of integrable lattice equations of BSQ-type. 
First, we showed that the nine-point lpBSQ equation is consistent around a $3\times 3 \times 3$ cube, \ie a cubic sublattice of $27$ vertices. This property generalises the standard consistency-around-a-cube property beyond the quadrilateral case.  Second, we constructed a new three-component quadrilateral system: the lattice BSQ–Q3 system, which serves as the BSQ analogue of the Q3 equation in the ABS classification \eqref{eq:fuvz}, or alternatively \eqref{eq:autouvz}.
The system is derived via a discrete gauge transformation between two Lax pairs of lpBSQ with distinct spectral parameters. In particular, there is  a free parameter $\delta$ appearing in BSQ-Q3 which arises from a $GL_3$ action on the lpBSQ Lax systems.  We proved that the system is 3D-consistent. A straight degeneration of BSQ-Q3 reduces to a $PGL_3$-invariant system \eqref{sl3-invar} generalising the well-known $PGL_2$-invariant Schwarzian BSQ equation.

These results provide concrete steps toward a systematic extension of integrable lattice equations beyond the KdV (ABS) framework. Several natural directions remain open for further investigation.
First, one may seek exact solutions of the lattice BSQ–Q3 system. Note that exact solutions of various lattice BSQ-type equations have been extensively studied (see, for example, \cite{MK, ZZN, NSZ,HZ2, HZ}), and one could extend the previous known solutions of lpBSQ to those of BSQ-Q3 via the eigenfunction expansions, \cf \eqref{eq:eig}. Second, it would be instructive to explore 
further degenerations of BSQ-Q3 leading to BSQ analogues of Q2 and Q1$_{\delta\neq0}$. This could be achieved by systematically degenerating both the equation and its exact solutions as was done in the KdV case \cite{NAH}.
Third, the  consistency around a $ 3 \times  3 \times   3$ property established here for lpBSQ may be extended to other nine-point BSQ-type equations, such as the lattice modified BSQ equation \cite{GD} and the lattice Schwarzian BSQ equation \eqref{eq:sl2bsq}. Fourth, it is worth noting that the $PGL_3$-invariant BSQ system \eqref{sl3-invar} together with the $PGL_3$-invariant BSQ equation  \eqref{eq:sl2bsq} differ from the multi-ratio systems considered, for instance, in \cite{KS, AJ},  which also claim to represent lattice BSQ-type equations invariant under certain projective transformations. It would be interesting  to  clarify the relationship between these formulations. Finally, one may aim to construct a BSQ counterpart of the elliptic Q4 equation, which would serve as a master equation in a higher-rank ABS-like classification

Our final remarks concern a possible geometric framework on the discrete gauge transformation approach used in this paper. As discussed in Remark \ref{rmk:gauge}, a continuous analogue of the gauge transformation method was developed in the context of KdV-type hierarchies, where it was interpreted as factorization of $SL_2$ loop group  \cite{GM}. The discrete gauge transformation presented here leading to lattice KdV-Q3 and BSQ-Q3 can be viewed as a discretized version of the loop group factorization method \cite{GM}, and we manage to extend it to the BSQ case (associated with $GL_3$). In particular, lpKdV and lpBSQ variables are connected to KdV-Q3 and BSQ-Q3 variables via respective Miura-type transformations.
It is well-established that the classical Miura chain connecting (continuous) KdV, modified KdV, and Schwarzian KdV admits a clear geometric interpretation in terms of (equivalence class of nondegenerate) projective curves in $P^1$ \cite{Wilson, MSTS} (via the correspondence between projective curves and the Sturm-Liouville operator \cite{OT}). It is natural to ask the question: what is the geometric framework of the Miura maps obtained from the loop group factorzation method. Some preliminary attempt suggests the maps are related to Grassmannians: the KdV–Q3 variable corresponds to combinations of Pl\"ucker coordinates on $Gr(2,4)$, and and the BSQ–Q3 triple $(u,v,z)$ can be build up from  Pl\"ucker coordinates $Gr(3,6)$. A systematic exploration of this geometric picture will be pursued in future work.

\medskip

\medskip

\noindent Pengyu Sun

\noindent {\rm Department of Mathematics, Shanghai University, Shanghai, 200444, China}

\noindent {\rm Newtouch Center for Mathematics of Shanghai University, Shanghai 200444, China}

\noindent \url{ sun8yu@shu.edu.cn}


~~~~~

\noindent Cheng Zhang

\noindent {\rm Department of Mathematics, Shanghai University, Shanghai, 200444, China}

\noindent {\rm Newtouch Center for Mathematics of Shanghai University, Shanghai 200444, China}

\noindent \url{ch.zhang.maths@gmail.com}

~~~~~

\noindent Frank Nijhoff

\noindent {\rm School of Mathematics, University of Leeds, Leeds LS2 9JT, United Kingdom}

\noindent  \url{f.w.nijhoff@leeds.ac.uk}

~~~~~

\end{document}